\documentclass[acmsmall]{acmart}

\AtBeginDocument{%
  }

\setcopyright{acmlicensed}
\acmJournal{PACMMOD}
\acmYear{2025} \acmVolume{3} \acmNumber{1 (SIGMOD)} \acmArticle{24} \acmMonth{2}\acmDOI{10.1145/3709674}

\received{July 2024}
\received[revised]{September 2024}
\received[accepted]{November 2024}

\usepackage{graphicx} % Required for inserting images
\usepackage{tikz}
\usepackage{listings}
\usepackage{multirow}
\usepackage{booktabs}
\usepackage{url} 
\usepackage{xspace}
\usepackage{balance}
\usepackage{listings}
\usepackage{float}
\usepackage{makecell}

\newcommand*\circled[1]{\tikz[baseline=(char.base)]{
            \node[shape=circle,draw,inner sep=0.1pt] (char) {#1};}}

\newcommand\redout{\bgroup\markoverwith
	{\textcolor{red}{\rule[.45ex]{1.5pt}{1.pt}}}\ULon}

 \newcommand\blueout{\bgroup\markoverwith
	{\textcolor{blue}{\rule[.45ex]{1.5pt}{1.pt}}}\ULon}

\definecolor{light-gray}{gray}{0.80}
\definecolor{ForestGreen}{RGB}{63,147,88}

\makeatletter
\newenvironment{btHighlight}[1][]
{\begingroup\tikzset{bt@Highlight@par/.style={#1}}\begin{lrbox}{\@tempboxa}}
	{\end{lrbox}\bt@HL@box[bt@Highlight@par]{\@tempboxa}\endgroup}

\newcommand\btHL[1][]{%
	\begin{btHighlight}[#1]\bgroup\aftergroup\bt@HL@endenv%
	}
	\def\bt@HL@endenv{%
	\end{btHighlight}%   
	\egroup
}
\newcommand{\bt@HL@box}[2][]{%
	\tikz[#1]{%
		\pgfpathrectangle{\pgfpoint{1pt}{0pt}}{\pgfpoint{\wd #2}{\ht #2}}%
		\pgfusepath{use as bounding box}%
		\node[anchor=base west, fill=orange!25,outer sep=.5pt,inner xsep=0.5pt, inner ysep=0.15pt, rounded corners=1pt, minimum height=\ht\strutbox-.1pt,#1]{\raisebox{.01pt}{\strut}\strut\usebox{#2}};
	}%
}
\makeatother

\lstdefinestyle{codestyle}{
	frame=none,
	commentstyle=\color{gray},
        columns=fixed, 
        keywordstyle=\color{black}\bfseries,
	numberstyle=\tiny\color{gray},
	stringstyle=\color{purple},
        basicstyle=\fontsize{7.9}{7.9}\ttfamily,
	breakatwhitespace=false,         
	breaklines=false,                 
	captionpos=t,                    
	keepspaces=true,     
	numbers=none,                    
	numbersep=4pt,                  
	showspaces=false,                
	showstringspaces=false,
	showtabs=false,                  
	tabsize=2,
	language=Sql,
	escapechar=|`,
        moredelim=**[is][{\btHL[fill=red!30]}]{@}{@},
        moredelim=**[is][{\btHL[fill=blue!30]}]{^}{^},
        morekeywords={with},
}
\definecolor{light-gray}{gray}{0.80}
\definecolor{ForestGreen}{RGB}{63,147,88}

\usepackage{soul}
\usetikzlibrary{calc}

\makeatletter
\newif\if@anonymize

\@anonymizefalse  % Uncomment to show text

\if@anonymize
  \newcommand{\highlight@DoHighlight}{
    \fill [outer sep = -15pt, inner sep = 0pt, color=black]
          ($(begin highlight)+(0,8pt)$) rectangle ($(end highlight)+(0,-3pt)$) ;
  }

  \newcommand{\highlight@BeginHighlight}{
    \coordinate (begin highlight) at (0,0) ;
  }

  \newcommand{\highlight@EndHighlight}{
    \coordinate (end highlight) at (0,0) ;
  }

  \newdimen\highlight@previous
  \newdimen\highlight@current
  \newlength{\item@width}

  \DeclareRobustCommand*\anonymize{%
    \SOUL@setup
    \def\SOUL@preamble{%
      \begin{tikzpicture}[overlay, remember picture]
        \highlight@BeginHighlight
        \highlight@EndHighlight
      \end{tikzpicture}%
    }%
    \def\SOUL@postamble{%
      \begin{tikzpicture}[overlay, remember picture]
        \highlight@EndHighlight
        \highlight@DoHighlight
      \end{tikzpicture}%
    }%
    \def\SOUL@everyhyphen{%
      \discretionary{%
        \SOUL@setkern\SOUL@hyphkern
        \SOUL@sethyphenchar
        \tikz[overlay, remember picture] \highlight@EndHighlight ;%
      }{%
      }{%
        \SOUL@setkern\SOUL@charkern
      }%
    }%
    \def\SOUL@everyexhyphen##1{%
      \SOUL@setkern\SOUL@hyphkern
      \settowidth{\item@width}{##1}%
      \makebox[\item@width]{}%
      \discretionary{%
        \tikz[overlay, remember picture] \highlight@EndHighlight ;%
      }{%
      }{%
        \SOUL@setkern\SOUL@charkern
      }%
    }%
    \def\SOUL@everysyllable{%
      \begin{tikzpicture}[overlay, remember picture]
        \path let \p0 = (begin highlight), \p1 = (0,0) in \pgfextra
          \global\highlight@previous=\y0
          \global\highlight@current =\y1
        \endpgfextra (0,0) ;
        \ifdim\highlight@current < \highlight@previous
          \highlight@DoHighlight
          \highlight@BeginHighlight
        \fi
      \end{tikzpicture}%
      \settowidth{\item@width}{\the\SOUL@syllable}%
      \makebox[\item@width]{}%
      \tikz[overlay, remember picture] \highlight@EndHighlight ;%
    }%
    \SOUL@
  }
\else
  \newcommand{\anonymize}[1]{#1}
\fi
\makeatother

\usepackage{pgfplots}
\usepackage{subcaption}

\usepackage{algpseudocode}

\usepackage{xcolor,amsmath}
\usepackage[linesnumbered,ruled,vlined]{algorithm2e}
\DontPrintSemicolon

\SetKwComment{Comment}{\color{green!50!black}// }{}

\newcommand{\assign}{\leftarrow}

\newcommand{\FuncCall}[2]{\texttt{ #1(#2)}}
\SetKwProg{Function}{function}{}{}

\newcommand{\eg}{\hbox{\emph{e.g.,}}\xspace}
\newcommand{\ie}{\hbox{\emph{i.e.,}}\xspace}
\title{Constant Optimization Driven Database System Testing}

\author{Chi Zhang}
\orcid{0000-0002-2848-6108}
\affiliation{
  \department{State Key Laboratory for Novel Software Technology, School of Computer Science}
  \institution{Nanjing University}
  \city{Nanjing}
  \country{China}
}
\email{zhangchi_seg@smail.nju.edu.cn}

\author{Manuel Rigger}
\orcid{0000-0001-8303-2099}
\affiliation{
  \institution{National University of Singapore}
  \country{Singapore}
}
\email{rigger@nus.edu.sg}

\newcommand*{\approach}{CODDTest\xspace}

\newcommand*{\totallogicbug}{24\xspace}
\newcommand*{\fixedlogicbug}{19\xspace}
\newcommand*{\conlogicbug}{5\xspace}

\newcommand*{\totalotherbug}{21\xspace}
\newcommand*{\fixedotherbug}{14\xspace}
\newcommand*{\conotherbug}{7\xspace}

\newcommand*{\totalbug}{45\xspace}
\newcommand*{\fixedbug}{33\xspace}
\newcommand*{\conbug}{12\xspace}

\begin{document}
% \date{}

\begin{abstract}
%Database systems play a crucial role in storing data for numerous essential software systems. 
Logic bugs are bugs that can cause database management systems (DBMSs) to silently produce incorrect results for given queries.
Such bugs are severe, because they can easily be overlooked by both developers and users, and can cause applications that rely on the DBMSs to malfunction.
% Several approaches have been proposed to detect logic bugs in DBMSs, but there is currently no general approach specifically designed to detect logic bugs in predicates, which are one of the most critical features of structured query language (SQL), and are used in multiple clauses.
In this work, we propose \textit{Constant-Optimization-Driven Database Testing} (\approach) as a novel approach for detecting logic bugs in DBMSs.
This method draws inspiration from two well-known optimizations in compilers: constant folding and constant propagation.
Our key insight is that for a certain database state and query containing a predicate, we can apply constant folding on the predicate by replacing an expression in the predicate with a constant, anticipating that the results of this predicate remain unchanged; any discrepancy indicates a bug in the DBMS.
We evaluated \approach on five mature and extensively-tested DBMSs--SQLite, MySQL, CockroachDB, DuckDB, and TiDB--and found \totalbug unique, previously unknown bugs in them.
Out of these, \totallogicbug are unique logic bugs. 
Our manual analysis of the state-of-the-art approaches indicates that 11 logic bugs are detectable only by \approach. 
% While \approach has lower test throughput compared to the state-of-the-art approaches, our evaluation indicates that the complexity of the queries generated by \approach is a major reason.
We believe that \approach is easy to implement, and can be widely adopted in practice. % understand and
\end{abstract}

\begin{CCSXML}
<ccs2012>
   <concept>
       <concept_id>10002951.10002952.10003190.10003192</concept_id>
       <concept_desc>Information systems~Database query processing</concept_desc>
       <concept_significance>500</concept_significance>
       </concept>
   <concept>
       <concept_id>10011007.10011074.10011099.10011102.10011103</concept_id>
       <concept_desc>Software and its engineering~Software testing and debugging</concept_desc>
       <concept_significance>500</concept_significance>
       </concept>
 </ccs2012>
\end{CCSXML}

\ccsdesc[500]{Information systems~Database query processing}
\ccsdesc[500]{Software and its engineering~Software testing and debugging}

\keywords{DBMSs testing, logic bugs, test oracle}

\maketitle

\section{Introduction}

Database management systems (DBMSs) store data for numerous essential software systems. %play a crucial role in storing data for numerous essential software systems. 
Similar to other software systems, DBMSs are susceptible to a range of bugs, including logic bugs. Logic bugs can be critical as they can cause the DBMSs to silently produce incorrect results for a given query and might be overlooked by both developers and users.

%Ensuring the correctness of DBMSs is of utmost importance. 
% Significant efforts have been made to detect logic bugs in DBMSs. 
In recent years, various approaches have been proposed to find logic bugs in DBMSs.
The state-of-the-art approaches are
NoREC~\cite{rigger2020detecting}, TLP~\cite{rigger2020finding}, PQS~\cite{rigger2020testing}, DQE~\cite{song2023testing}, and TQS~\cite{DBLP:journals/pacmmod/TangWZ0023}. 
NoREC and DQE assume that the same predicate should, for a given row, consistently evaluate to the same value,
regardless of which clause it is used in. 
NoREC applies this to \lstinline[style=codestyle]{WHERE} clauses in \lstinline[style=codestyle]{SELECT}, while DQE applies this to \lstinline[style=codestyle]{WHERE} clauses in \lstinline[style=codestyle]{UPDATE} and \lstinline[style=codestyle]{DELETE}.
TLP decomposes a query into three partitioning queries, each retrieving rows based on predicates \lstinline[style=codestyle]{p}, \lstinline[style=codestyle]{NOT p}, and \lstinline[style=codestyle]{p IS NULL}, respectively. 
All these approaches are black-box techniques, as they apply on the SQL level, not on the DBMSs' source code.
Because of this, and since automated testing approaches typically lack guarantees, it is difficult to precisely characterize what bugs they find and which ones they miss.
For one, none of these approaches support testing subqueries, an important feature that allows query nesting, and it is not obvious how these approaches could be extended to support it.
For another, as the results of this paper show, these approaches overlook various bugs due to their inherent limitations.

To tackle logic bugs in DBMSs, we propose a novel testing methodology that tests DBMSs through
% We tackle this issue through an interdisciplinary stance by testing database systems through 
the lens of constant folding and constant propagation, two optimizations that were originally proposed for compilers.
Specifically, we propose \textit{Constant-Optimization-Driven Database Testing (CODDTest)},\footnote{This name pays homage to Edgar Frank Codd~\cite{DBLP:journals/cacm/Codd70}, who proposed the relational model for database management.} which is a black-box approach that, for a given database state and query, applies constant folding and propagation to expressions in the query.
The \emph{test oracle}, that is, the mechanism for validating the behavior of the DBMS for a given input, relies on validating that the transformed query produces the same result as the original query.

Constant folding is a well-known compiler optimization that evaluates constant expressions at compile time, rather than computing them at run time~\cite{DBLP:books/mk/Muchnick1997}.
For example, it evaluates the statement \lstinline[style=codestyle]{i = 1 + 2 + 3;} to \lstinline[style=codestyle]{i = 6;} at compile time.
Constant folding extends beyond numeric literal expressions by leveraging constant propagation.
Constant propagation, through reachability analysis~\cite{chen1998c++}, determines constant values for variables by assessing their reachability at specific program points.
By employing constant propagation in conjunction with constant folding, more intricate programs can be optimized. For example, the statement sequence \lstinline[style=codestyle]{a = 1; i = a + 2 + 3;} can be optimized to \lstinline[style=codestyle]{i = 6;}.

In the context of DBMSs, % applying constant folding and constant propagation is easier than in compilers, as reachability analysis is not required.
the result of an expression in a query is deterministic for any given row under a specific database state.
While, as an optimization, applying constant folding and propagation assuming a constant database state would be ineffective, since the state typically frequently changes, we can leverage these optimizations for a test oracle.
Listing~\ref{fig:illustexample} illustrates this by using a bug-inducing test case that enabled us to find a subquery-related bug in SQLite.\footnote{\url{https://sqlite.org/forum/info/\anonymize{a68313d0545273c8}}}
We start from an initial database state, which includes a table \lstinline[style=codestyle]{t0} with index \lstinline[style=codestyle]{i0}, as well as a view \lstinline[style=codestyle]{v0}.
%Table \lstinline[style=codestyle]{t0} contains only one row (\ie \lstinline[style=codestyle]{1}), and we can use this value in the \lstinline[style=codestyle]{CREATE VIEW} statement to ascertain that there is also a single row (\ie \lstinline[style=codestyle]{1}) within \lstinline[style=codestyle]{v0}.
The query \circled{O}, which corresponds to the \emph{original query}, uses a subquery in the \lstinline[style=codestyle]{WHERE} clause, which we want to constant-fold.
We can do so by extracting the subquery, using an \emph{auxiliary query}, as shown in query \circled{A}, and executing it using the DBMS under test, which returns~\lstinline[style=codestyle]{0}.
%In the subquery, \lstinline[style=codestyle]{v0.c0} contains only one row (\ie \lstinline[style=codestyle]{1}), and the predicate \lstinline[style=codestyle]{v0.c0 BETWEEN 0 AND 0} evaluates to false on it. Consequently, the result of the subquery should be \lstinline[style=codestyle]{0}, since no rows are fetched.
%Therefore, by considering the subquery, which consistently evaluates to \lstinline[style=codestyle]{0}, as the predicate for the outer query, we anticipate that the result of query \circled{1} will be \lstinline[style=codestyle]{0}.
We derive the \emph{folded query} as shown in query \circled{F}, by, in query \circled{O}, replacing the subquery with its result obtained in query \circled{A}.
Unexpectedly, SQLite returned a different result for the original and folded queries; for query \circled{O}, SQLite returned \lstinline[style=codestyle]{1}, while it returned \lstinline[style=codestyle]{0} for query \circled{F}.
This discrepancy indicates a bug in the DBMS.
%Unexpectedly, upon deriving its result (\ie \lstinline[style=codestyle]{1}) from the DBMS under test, we discovered that it is incorrect. 
%We found this bug by applying constant folding on the subquery of the query \circled{1}. 
%We extract the subquery into a single query (query \circled{2}) and retrieve its results (\ie \lstinline[style=codestyle]{0}) from the DBMS under test based on the current state of the database.
%Last, we replace the subquery of query \circled{1} with its result to generate the query \circled{3}, and derive the result (\ie \lstinline[style=codestyle]{0}), which is different from the result of query \circled{1}. The discrepancy between these two queries indicates that the DBMS is affected by a bug. 
We received feedback from the developers of SQLite, stating that this bug was caused by a query planner optimization, and an aggregate subquery is a necessary condition to trigger this bug. We found this bug, since our folded query (\ie query \circled{F}) no longer met this requirement.
Our approach is akin to a combination of constant folding and constant propagation.
In this example, using the DBMS to evaluate the expression under a certain database state can be seen as constant folding (\ie in query \circled{A}), while the substitution of the entire expression with its result can be viewed as constant propagation (\ie replacing the subquery in query \circled{O} with the constant \lstinline[style=codestyle]{0}).

\begin{lstlisting}[style=codestyle, belowskip=0pt, caption={An illustrative example, which triggered a bug in SQLite. The expression in the original query subject to constant folding and propagation, along with the corresponding constant in the folded query, are highlighted in red.}, label=fig:illustexample, float, floatplacement={tb}]
CREATE TABLE t0 (c0);
INSERT INTO t0(c0) VALUES (1);
CREATE INDEX i0 ON t0(c0 > 0);
CREATE VIEW v0(c0) AS SELECT AVG(t0.c0) FROM t0 GROUP BY 1 > t0.c0;
|`\circled{O}|` SELECT COUNT(*) FROM t0 INDEXED BY i0 WHERE (@SELECT COUNT(*) FROM v0 WHERE v0.c0@ 
    @BETWEEN 0 AND 0@); -- 1 |`$\vcenter{\hbox{\includegraphics[height=\baselineskip]{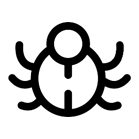}}}$|`
|`\circled{A}|` SELECT COUNT(*) FROM v0 WHERE v0.c0 BETWEEN 0 AND 0; -- 0
|`\circled{F}|` SELECT COUNT(*) FROM t0 INDEXED BY i0 WHERE @0@; -- 0 |`$\vcenter{\hbox{\includegraphics[height=\baselineskip]{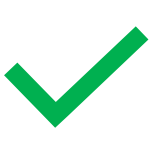}}}$|`
\end{lstlisting}

Our approach is effective in detecting logic bugs, because it can directly transform a predicate, resulting in the execution of different code within DBMSs, while the folded query should produce the same results as the original one.
Any discrepancy between the queries indicates a bug in DBMSs. 
Our approach is capable of identifying bugs that occur simultaneously in all three \lstinline[style=codestyle]{SELECT}, \lstinline[style=codestyle]{UPDATE}, and \lstinline[style=codestyle]{DELETE} statements,
% By not relying on the fetched rows to determine predicate correctness, we detect bugs in clauses that do not directly impact the results, 
which will be missed by NoREC and DQE.
Our approach can detect bugs where boolean predicates consistently evaluate to an opposite value, which would be missed by TLP.
Our transformation-based oracle can handle complex queries, whereas PQS is limited by the ability of the self-implemented interpreter.
Overall, we believe that \approach complements existing testing approaches for DBMSs, and provides a new conceptual angle on using compiler optimizations for DBMS testing.
%Overall, we thus believe that \approach, despite its simplicity, is effective in detecting various bugs that have been missed by existing approaches.

% Specifically, our approach allows us to prevent multiple optimizations on predicates, such as predicate move-around~\cite{levy1997query} and magic-sets rewriting~\cite{mumick1990magic}. 
% Additionally, our approach tests operators and functions when they receive expressions and constants as arguments.

% We implemented \approach based on SQLancer, 
We integrated \approach into SQLancer,\footnote{\url{https://github.com/sqlancer/sqlancer/pull/1054}}
a widely-used tool for testing DBMSs, in which also other state-of-the-art oracles have been integrated, and tested five mature  DBMSs with it, all of which have been extensively tested by a number of state-of-the-art testing approaches. 
The results were surprisingly positive. We found \totalbug previously unknown, unique bugs. Of these, \conbug were confirmed, \fixedbug were fixed.
% , and none of them is waiting for confirmation.
\totallogicbug out of these were previously unknown unique logic bugs, which we aimed to find. Of these logic bugs, \conlogicbug were confirmed, and \fixedlogicbug were fixed. 
%All of these systems have already been extensively tested by a number of state-of-the-art testing approaches.
In summary, we make the following contributions:
\begin{itemize}
  \item We propose the novel idea of testing DBMSs through the lens of constant propagation and constant folding.
  \item We provide a realization of this idea by, given an \emph{original query}, deriving a \emph{folded query}, in which an expression has been constant-folded based on the result of an \emph{auxiliary query} to validate whether the DBMS computes a consistent result for the original and folded queries.
  \item We implemented the approach and evaluated it on five well-tested and mature DBMSs, uncovering \totallogicbug unique, previously unknown logic bugs. Furthermore, we provide a comparative analysis with the state-of-the-art approaches, as well as a performance comparison.
\end{itemize}

\section{Background}
% \paragraph{Relational DBMSs.}
\paragraph{Predicates}
% The relational model~\cite{DBLP:journals/cacm/Codd70} is the predominant model for DBMSs, which we consider in this paper. 
% In relational DBMSs, tables consist of columns, which represent data attributes, and each row in the table represents an instance of the relation. 
% Consequently, each row in the table should yield its own result for a given expression.
% Structured query language (SQL) is the programming language that applications and users can use to interact with the DBMS.
In SQL, a predicate is a boolean expression that evaluates to \lstinline[style=codestyle]{TRUE}, \lstinline[style=codestyle]{FALSE}, or \lstinline[style=codestyle]{NULL} when applied to given values or rows.
Predicates are used in various clauses of SQL, such as the \lstinline[style=codestyle]{WHERE} clauses of \lstinline[style=codestyle]{SELECT}, \lstinline[style=codestyle]{UPDATE}, and \lstinline[style=codestyle]{DELETE}, as well as the \lstinline[style=codestyle]{JOIN ON}, \lstinline[style=codestyle]{HAVING}, \lstinline[style=codestyle]{GROUP BY}, and \lstinline[style=codestyle]{ORDER BY} clauses of \lstinline[style=codestyle]{SELECT}. 

\begin{lstlisting}[style=codestyle, belowskip=0pt, caption={A correlated subquery example.}, label={fig:correlated}, float, floatplacement={tb}]
CREATE TABLE t0(ID INT, score INT, classID INT);
INSERT INTO t0 VALUES (0, 90, 1), (1, 80, 1), (2, 83, 2);
|`\circled{O}|` SELECT x.ID FROM t0 AS x WHERE x.score > 
    (@SELECT AVG(y.score) FROM t0 AS y WHERE x.classID = y.classID@);
|`\circled{A}|` SELECT x.classID, 
    (SELECT AVG(y.score) FROM t0 AS y WHERE x.classID = y.classID) FROM t0 AS x;
|`\circled{F}|` SELECT x.ID FROM t0 AS x WHERE x.score > 
    (@CASE WHEN x.classID = 1 THEN 85@ 
          @WHEN x.classID = 1 THEN 85@ 
          @WHEN x.classID = 2 THEN 83 END@);
\end{lstlisting}
\begin{lstlisting}[style=codestyle, belowskip=0pt, caption={A {\lstinline[style=codestyle]{CASE}} expression example.}, label=fig:caseexample, float, floatplacement={tb}]
CREATE TABLE grade (score INT);
INSERT INTO grade(score) VALUES (100), (80), (60);
SELECT score, CASE WHEN score = 100 THEN "A"
                   WHEN score >= 80 AND score < 100 THEN "B"
                   ELSE "C" END FROM grade;
\end{lstlisting}

\paragraph{Subqueries.}
Subqueries are an important language feature in SQL, and can be used in predicates.
Subqueries can be classified as correlated or non-correlated based on whether the subquery references columns from the outer query.
Correlated subqueries are \lstinline[style=codestyle]{SELECT} queries nested within outer queries, referencing columns from the outer queries to construct their predicate. A correlated subquery will be executed once for each row passed from the outer query.
Consider the query \circled{O} depicted in Listing~\ref{fig:correlated}, where a correlated subquery is used to identify students (\ie represented by \lstinline[style=codestyle]{ID}) whose scores exceed the average score in the class.
The subquery calculates the average score of the class, whose \lstinline[style=codestyle]{classID} is passed from the outer query. 
Therefore, for each student in \lstinline[style=codestyle]{t0}, the subquery will be executed once to derive the average score in the class of that particular student. 
In contrast, non-correlated subqueries do not reference columns in the outer query and are evaluated once in execution.
Various optimizations have been developed to enhance the performance of subqueries, such as subquery unnesting~\cite{DBLP:conf/icde/Celis97, DBLP:conf/icde/BrantnerMM07} or correlated subquery decorrelation~\cite{DBLP:conf/icde/SeshadriPL96}. 
Like other language features, subquery support in DBMSs is susceptible to logic bugs. 
% However, existing state-of-the-art logic bug detection approaches for DBMSs lack comprehensive testing for subqueries.

\paragraph{SQL \lstinline[style=codestyle]{CASE} expression.}

The \lstinline[style=codestyle]{CASE} expression is a feature of SQL to process \lstinline[style=codestyle, morekeywords={if}]{if/then} logic.
Listing~\ref{fig:caseexample} illustrates an example using the \lstinline[style=codestyle]{CASE} expression. The predicate following the \lstinline[style=codestyle]{WHEN} keyword determines whether the value after the \lstinline[style=codestyle]{THEN} keyword should be returned. If all predicates evaluate to false, the result of this \lstinline[style=codestyle]{CASE} expression is the value specified in the \lstinline[style=codestyle]{ELSE} clause, or \lstinline[style=codestyle]{NULL} if there is no \lstinline[style=codestyle]{ELSE} clause.
In our approach, we use the \lstinline[style=codestyle]{CASE} expression to map each row of a table to constant values, corresponding to constant folding on a per-row basis.

\paragraph{Metamorphic testing.}
Metamorphic testing~\cite{DBLP:journals/corr/abs-2002-12543} is an automated testing methodology, which generates a new input for the program under test, based on an existing input and its corresponding output, with the expectation that the outcome of this new input can be predicted.
Formally, given an input $I$ and $P(I) = O$, where $P$ is the program under test, a follow-up input $I'$ is derived, so that a known relationship between $O$ and $P(I') = O'$ is validated.
Metamorphic testing has been applied in testing numerous critical systems, such as compilers~\cite{le2014compiler}, SMT solvers~\cite{DBLP:conf/pldi/WintererZS20}, and DBMSs.
NoREC~\cite{rigger2020detecting}, TLP~\cite{rigger2020finding}, and DQE~\cite{song2023testing} are all metamorphic testing approaches for detecting logic bugs in DBMSs. 
The core challenge of realizing a metamorphic testing approach is to identify a so-called metamorphic relation that  derives the follow-up input $I'$ and relates the outputs $O$ and $O'$ so that bugs in the system under test can be revealed.

\section{Approach}
\label{sect:approach}

We propose \approach as an approach to finding logic bugs in DBMSs through the lens of constant folding and propagation.
Our key insight is that within an SQL query, by assuming a constant database state and given query, we can apply constant folding and constant propagation to a specific expression in a predicate, assuming that the query's result remains unchanged. 
More precisely, our approach for testing DBMSs generates two equivalent queries:  the \textit{original query}, which includes the randomly generated predicate, and the \textit{folded query}, which is derived by substituting the predicate in the original query with the corresponding constant-folded predicate.
Any discrepancy in the results of these two queries indicates a potential bug.

\paragraph{Metamorphic relation.}
\approach is a metamorphic testing approach, where the folded query $F$ is derived from the original query $O$, and the results of these two queries with DBMS engine $E$ are expected to be identical.
More formally, for the potential database state space denoted by $\mathcal{S}$, we can establish the metamorphic relationship below. Any violation indicates a bug in $E$.
\begin{equation}\nonumber
% \label{eq:mtrel}
\forall s \in \mathcal{S}\quad E_s(O) = E_s(F)
\end{equation}
where $E_s(O)$ denotes the outcome of executing a query $O$ with the DBMS engine $E$ under the state $s$, while $E_s(F)$ represents the result of $F$.

The generation of $F$ involves applying constant folding and constant propagation to a randomly selected expression $\phi$ within the predicate of $O$, represented as $F = O\lbrack \phi/R_{\phi}\rbrack$, which includes substituting $\phi$ with the expression $R_{\phi}$ within the same query location in the query $O$. 
$R_{\phi}$ is the evaluated result of $\phi$, and this process referred to as constant propagation.
$R_{\phi}$ can be obtained through constant folding, which entails evaluating $\phi$ under the database $s$, denoted by $R_{\phi}=E_s(A\lbrack\phi\rbrack)$. Here, $A\lbrack\phi\rbrack$ represents an auxiliary query $A$ designed to obtain the evaluation result of $\phi$.
Finally, the above metamorphic relation can be elaborated as follows:

\begin{align*}
    \forall s \in \mathcal{S}\quad\quad\quad R_{\phi} &= E_s(A\lbrack\phi\rbrack)\\
    E_s(O) &= E_s(F) = E_s(O\lbrack \phi/R_{\phi}\rbrack) 
\end{align*}

\begin{figure*}[bt]
    \centering
    \includegraphics[width=.99\linewidth]{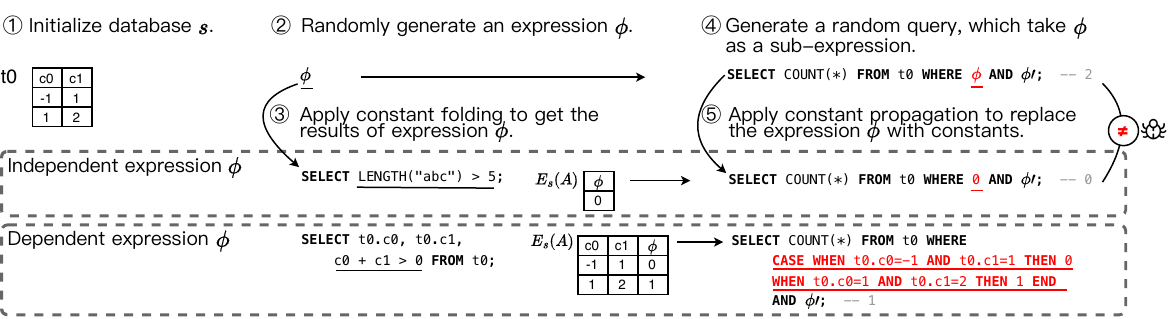}
    \vspace{-7pt}
    \caption{Overview of approach. \approach generates pairs of equivalent queries by applying constant folding and propagation to the expression $\phi$. The application of constant propagation and folding differs for independent and dependent expressions.}
    \vspace{-7pt}
    \label{fig:approach}
\end{figure*}

\sloppy{}
\paragraph{Approach overview.}
Figure~\ref{fig:approach} illustrates our approach. 
In step \circled{1}, we initialize the database and create non-empty tables. 
We do this randomly by using rule-based generators, such as also done by other testing works~\cite{rigger2020detecting, sqlsmith}.
Non-empty tables ensure that at least one row is available for us to apply constant folding.
Here, we generate a table \lstinline[style=codestyle]{t0} with two rows.
Subsequently, we enter a loop from step \circled{2} to step \circled{5} to thoroughly test the generated database state. Each iteration is designed to test the DBMS once. 
Step \circled{2} generates a random expression $\phi$, which will undergo constant folding.  
In step \circled{3}, we introduce the \textit{auxiliary query} to retrieve the corresponding constants of $\phi$ with respect to the database state---we see this step as constant folding.
In step \circled{4}, we generate the original query, using $\phi$ as part of a predicate.
We execute the original query using the DBMS under test and obtain its results (\ie~\lstinline[style=codestyle]{2}).
In step \circled{5}, we apply constant propagation to $\phi$ by replacing it with the corresponding constants to obtain the folded query. 
If the folded query produces a different result than the original query, we have found a bug-inducing test case.

Applying constant folding (\ie step \circled{3}) and constant propagation (\ie step \circled{5}) differs based on whether the expression $\phi$ is a so-called \emph{independent expression} or \emph{dependent expression}. 
Independent expressions yield constant results irrespective of the outer context, allowing us to perform constant folding and propagation on the expression $\phi$ by replacing it with a constant or a constant list.
For example, the independent expression $\phi$ shown in Figure~\ref{fig:approach}, \lstinline[style=codestyle]{LENGTH("abc") > 5}, does not reference any columns and evaluates to \lstinline[style=codestyle]{FALSE}, regardless of the database state and the clauses it is used in.
Similarly, the constant-folded expression (\ie query \circled{A}) in Listing~\ref{fig:illustexample} is an independent expression, as the non-correlated subquery can be executed independently and yields constant results regardless of the outer query.
We categorize expressions as dependent expressions if they reference columns from an outer context and cannot be executed independently. 
For example, the dependent expression \lstinline[style=codestyle]{c0 + c1 > 0} shown in Figure~\ref{fig:approach} references columns \lstinline[style=codestyle]{c0} and \lstinline[style=codestyle]{c1} from the \lstinline[style=codestyle]{FROM} clause; the original query shown in Listing~\ref{fig:correlated} is also a dependent expression, as the correlated subquery depends on the value of \lstinline[style=codestyle]{x.classID} from the outer query.
For dependent expressions, we must consider that $\phi$ might evaluate to a different value for every row, which is why we represent the folded constant as a mapping from a row value to the constant. 

\begin{algorithm}[tb]
\footnotesize
\caption{Algorithm of \approach}
\label{alg:oraclealg}

\Function{TestOracleGen($DatabaseState\ s$)}{
    \Comment{Randomly generate an expression $\phi$, which will undergo constant folding and constant propagation. We extract the set of the referenced columns $\{c_i\}$ in $\phi$, which come from outer context, along with the tables set $\{t_i\}$ to which these columns $c_i$ are associated.}
    $\phi, \{c_i\}, \{t_i\} \assign$  \FuncCall{GenExpr}{$s$}\;\label{alg:phigen}
    
    \Comment{Constant folding of $\phi$ under $s$, this step differs based on $\phi$ is dependent or independent expressions. Specifically, $\phi$ is considered an independent expression when $\{c_i\}$ is empty; otherwise, $\phi$ is classified as a dependent expression.}
    \If{\FuncCall{Size}{$\{c_i\}$} == 0}{\label{alg:cpbegin}
        \Comment{Construct the auxiliary query for independent expression}
        $A \assign$ "\textcolor{purple}{SELECT} " + $\phi$\;\label{alg:indexp}
        $E_s(A) \assign$ \FuncCall{ExecQuery}{$A$, $s$}\;\label{alg:resind}
        \Comment{Transform the constant result of $\phi$ as a constant expression $R_{\phi}$}
        $R_{\phi} \assign E_s(A)$\;\label{alg:constexp}
    }
    \Else{
        \Comment{Dependent expression has different result for each row of $\{c_i\}$}
        $A \assign$ "\textcolor{purple}{SELECT} " + $(\{c_i\}, \phi)$ + "\textcolor{purple}{FROM} " + $\{t_i\}$\;\label{alg:dexp}
        $E_s(A) \assign$ \FuncCall{ExecQuery}{$A$, $s$}\;
        \Comment{Map the results of $\phi$ to each row of $\{c_i\}$ as an expression $R_{\phi}$}
        $R_{\phi} \assign$ \FuncCall{Map}{$E_s(A)$}\;\label{alg:cpend}
    }
    \Comment{Generate the original query based on the current database state, using $\phi$ as a sub-expression in predicate}
    $O \assign$ \FuncCall{QueryGenerate}{$s$, $\phi$, $\{t_i\}$}\;\label{alg:orggen}
    $E_s(O) \assign$ \FuncCall{ExecQuery}{$O$, $s$}\;\label{alg:orgres}
    \Comment{Generate the folded query by replacing $\phi$ with $R_{\phi}$}
    $F \assign$ \FuncCall{ReplaceExpr}{$\phi$, $R_{\phi}$, $O$}\;\label{alg:foldgen}
    $E_s(F) \assign$ \FuncCall{ExecQuery}{$F$, $s$}\;\label{alg:folres}
    \Comment{A bug is identified if there is a discrepancy between the results of the original query and the folded query}
    \If{$E_s(O) != E_s(F)$}{\label{alg:bug1}
        \FuncCall{ReportBug}{$O$, $F$, $A$}\;\label{alg:bug2}
    }
}
\end{algorithm}

\paragraph{Algorithmic sketch.}
We use Algorithm~\ref{alg:oraclealg} to demonstrate the process of generating a test oracle.
The function \texttt{TestOracleGen} corresponds to step~\circled{2} to step~\circled{5} in Figure~\ref{fig:approach}, taking the randomly generated database state $s$ as its input.

For step~\circled{2}, we randomly generate the expression $\phi$ with function \texttt{GenExpr}, based on the current database state $s$, as outlined in line~\ref{alg:phigen}. $\phi$ then undergoes constant folding and constant propagation.
\texttt{GenExpr} returns two sets: $\{c_i\}$, which consists of referenced columns from the outer context and is utilized to determine if $\phi$ is an independent expression, and $\{t_i\}$, which denotes the tables to which $c_i$ belong.
For example, for the independent expression \lstinline[style=codestyle]{LENGTH("abc") > 5}, \texttt{GenExpr} yields two empty sets for $\{c_i\}$ and $\{t_i\}$ as results, due to the absence of any referenced columns in this expression.  
Conversely, for the dependent expression \lstinline[style=codestyle]{c0 + c1 > 0} shown in Figure~\ref{fig:approach}, \texttt{GenExpr} returns $\{c0, c1\}$, and $\{t0\}$.

The logic for constant folding (\ie step~\circled{3}) spans from line~\ref{alg:cpbegin} to line~\ref{alg:cpend}.
When the size of $\{c_i\}$ is zero, $\phi$ does not reference any columns from the outer context, leading to $\phi$ producing constant results regardless of the outer context. Therefore, $\phi$ is considered an independent expression.
The value of an independent expression can be obtained through an auxiliary query consisting solely of the \lstinline[style=codestyle]{SELECT} keyword followed by $\phi$, as shown in line~\ref{alg:indexp}. 
Moreover, this \lstinline[style=codestyle]{SELECT} keyword can be omitted when $\phi$ is a non-correlated subquery.
The outcome of an independent expression is either a constant (the empty result can be considered as \lstinline[style=codestyle]{NULL}) or a constant list. Once we ascertain the result of the auxiliary query (\ie line~\ref{alg:resind}), we can directly convert this result into a constant expression in line~\ref{alg:constexp}.
For a dependent expression, which produces different outcomes for various rows of $\{c_i\}$, we must construct $R_{\phi}$ to reflect this variability.
In line~\ref{alg:dexp}, by placing $\{c_i\}$ and $\phi$ together following the \lstinline[style=codestyle]{SELECT} keyword, we obtain the result of $\phi$, as well as the values of $\{c_i\}$ that lead to the result of $\phi$.
To ensure syntactic correctness, we also need to append the list of tables $\{t_i\}$ after the \lstinline[style=codestyle]{FROM} keyword.
Finally, we construct $R_{\phi}$ as a mapping from the value of each row in $\{c_i\}$ to the corresponding result of $\phi$, as shown in line~\ref{alg:cpend}.
In the subsequent two subsections, we elaborate and further illustrate how we handle both independent and dependent expressions.

Step~\circled{4} is accomplished in line~\ref{alg:orggen} through the \texttt{QueryGenerate} function, a random query generator that takes the current database state $s$, the expression $\phi$, and the table set $\{t_i\}$ as input. $\phi$ is then randomly incorporated into a predicate of the generated query.
It is imperative to include the table set $\{t_i\}$ as an input. This requirement arises from the need for auxiliary queries to replicate the original query's \lstinline[style=codestyle]{JOIN} clauses, with the sole exception being instances where $\phi$ functions as the predicate within a \lstinline[style=codestyle]{JOIN} clause. 
The set $\{t_i\}$ records the information of the \lstinline[style=codestyle]{JOIN} clauses associated with each table. 
For a deeper understanding of the reason, we provide the explanation in Section~\ref{sec:foldde}, in which we discuss how to perform constant folding on dependent expressions.

Constant propagation (\ie step~\circled{5}) is applied in line~\ref{alg:foldgen} using the \texttt{ReplaceExpr} function, which replaces the expression $\phi$ with the expression $R_{\phi}$---representing the result of $\phi$, in the corresponding location within query $O$.
After obtaining the result of the original query $O$ in line~\ref{alg:orgres} and the result of the folded query in line~\ref{alg:folres}, we can identify the presence of a bug if any discrepancies exist between these two results (\ie line~\ref{alg:bug1} to line~\ref{alg:bug2}).

\subsection{Folding Independent Expressions}
Independent expressions can be executed without surrounding contexts and can have two possible shapes.
First, if the expression $\phi$ has no column references, it is a constant expression that always yields a constant value. In such cases, in the auxiliary query, we can evaluate the expression using a \lstinline[style=codestyle]{SELECT} statement. %, without the need to construct a mapping.
For example, the independent expression \lstinline[style=codestyle]{LENGTH("abc") > 5} shown in Figure~\ref{fig:approach} is used in a \lstinline[style=codestyle]{SELECT} statement to derive its results.
Second, the expression  $\phi$ can be a non-correlated subquery, which computes a constant result assuming a fixed database state.
%Under a given database state, a non-correlated subquery can be considered as a specific case of a constant expression, as the non-correlated subquery is independent of the outer query and consistently returns the same results.
%As a result, we can directly represent the non-correlated subquery using its corresponding results without a mapping.
For example, in Listing~\ref{fig:illustexample}, the subquery of query \circled{O} returns the same result regardless of the outer query's result.
Therefore, in the auxiliary query \circled{A}, we can directly execute the extracted subquery.
For both shapes, we create the folded query by replacing the expression with its evaluated result.
For example, in Listing~\ref{fig:illustexample}, \circled{A} evaluates to \lstinline[style=codestyle]{0}, which is why we replace the subquery with \lstinline[style=codestyle]{0} in \circled{F}.
%apply our approach by substituting the subquery in query \circled{O} with the value \lstinline[style=codestyle]{0}, resulting in query \circled{F}.
%Therefore, we can directly apply our approach by substituting the subquery in query \circled{O} with the value \lstinline[style=codestyle]{0}, resulting in query \circled{F}.
%However, a difference is that while a constant expression always returns a scalar value, 
%the results of a subquery can have different size: a scalar value, a row value,\footnote{https://www.sqlite.org/rowvalue.html} or multiple row value.
%To ensure semantic correctness, we will introduce how to construct the predicate based on the shape of the expression $\phi$'s results, and use the predicate in which clauses in Section~\ref{sect:queryconstruct}. 

\subsection{Folding Dependent Expressions}
\label{sec:foldde}

We conceptualize a dependent expression as a function $R_{\phi}=F(\theta), \theta=\{c_1, c_2, ..., c_i\}$, where the dependent expression can be seen as a mapping from its arguments $c_1, c_2, ..., c_i$ to its results $R_{\phi}$. 
Each argument $c_i$ is referenced in $\phi$ and represents a column from the outer context, and its domain consists of the rows within that column of the table. 
During constant folding and propagation, we must consider that $\phi$ might evaluate to a different value for every table row.
We first obtain the results of the expression on each row (\ie step \circled{3} for dependent expression) and then represent them using a mapping (\ie step \circled{5} for dependent expression).

%\approach applies constant propagation and folding for each row of the columns that come from outer context and are referenced by the dependent expression.
%Thus, for dependent expressions, we derive a map from each row of the columns that $\phi$ depends on to the results of $\phi$, as each row of the columns should yield its own result for a given expression.
%We use the \lstinline[style=codestyle]{CASE} expression to represent the results of $\phi$ for each row of the columns that $\phi$ depends on in the folded query.

\paragraph{Constant folding.}
To construct the auxiliary query in step \circled{3} for dependent expressions, we first identify $\theta$, which serves as the keys of the mapping, by collecting all the columns that appear in $\phi$ and reference the outer context.
We include these columns in the fetch clause.
Additionally, we include $\phi$ itself in the fetch clause to obtain its results for each row, which corresponds to $R_{\phi}$.
Each row of the auxiliary query's result contains the values of $\theta$, as well as the values of $R_{\phi}$, which are the results of $\phi$.
Therefore, each row of the results represents a map entry, where the values of $\theta$ serve as keys and the values of $R_{\phi}$ as corresponding map values.
For example, the dependent expression \lstinline[style=codestyle]{c0 + c1 > 0} shown in Figure~\ref{fig:approach} references two columns: \lstinline[style=codestyle]{c0} and \lstinline[style=codestyle]{c1}. These two columns as well as the expression itself are included in the fetch clause of the auxiliary query.
The auxiliary query's results consist of two rows, each representing the result of $\phi$ for the corresponding row of the table.
As another example, Listing~\ref{fig:correlated} shows how we create the auxiliary query for correlated subqueries.
To obtain the results of the subquery of query \circled{O} on all rows passed from the outer query, we include the subquery in the fetch clause as shown in query \circled{A}.
While the subquery references two columns \lstinline[style=codestyle]{x.classID} and \lstinline[style=codestyle]{y.classID} in the predicate, only \lstinline[style=codestyle]{x.classID} is from the outer context. Thus, in query \circled{A}, $\theta$ contains only \lstinline[style=codestyle]{x.classID}.

\begin{lstlisting}[style=codestyle, belowskip=-7pt, caption={{\lstinline[style=codestyle]{JOIN}} can affect the values of {$\phi$} for a given row.}, label={fig:joinabsence}, float, floatplacement={tb}]
CREATE TABLE t0 (c0 INT);
CREATE TABLE t1 (c0 INT);
INSERT INTO t0 VALUES (0);
INSERT INTO t1 VALUES (1);
|`\circled{O}|` SELECT * FROM t0 LEFT JOIN t1 ON t0.c0 = t1.c0 WHERE @t1.c0 IS NULL@; -- 0|NULL
|`\circled{A}|` SELECT t1.c0, t1.c0 IS NULL FROM t0 LEFT JOIN t1 ON t0.c0 = t1.c0; --NULL|1
|`\circled{F}|` SELECT * FROM t0 LEFT JOIN t1 ON t0.c0 = t1.c0 WHERE 
    @CASE WHEN t1.c0 is NULL THEN 1 END@; --0|NULL 
\end{lstlisting}

Supporting \lstinline[style=codestyle]{JOIN}s involves two considerations.
First, the auxiliary queries must use the same \lstinline[style=codestyle]{JOIN} clauses as the original query, except in cases where $\phi$ serves as the predicate within the \lstinline[style=codestyle]{JOIN} clause.
Listing~\ref{fig:joinabsence} shows an example where we assign $\phi$ the concrete expression \lstinline[style=codestyle]{t1.c0 IS NULL}.
The original query fetches a single row whose \lstinline[style=codestyle]{t1.c0} column holds a \lstinline[style=codestyle]{NULL} value. This is because the predicate of the \lstinline[style=codestyle]{LEFT JOIN} evaluates to false. Then, the expression \lstinline[style=codestyle]{t1.c0 IS NULL} is evaluated to be true. As a result, the mapping assigns only one row in \lstinline[style=codestyle]{t1.c0} to the result of \lstinline[style=codestyle]{t1.c0 IS NULL}: from \lstinline[style=codestyle]{NULL} to \lstinline[style=codestyle]{TRUE}. To construct this mapping in the auxiliary query,  the same \lstinline[style=codestyle]{JOIN} as the original query must be used, ensuring that the only row in \lstinline[style=codestyle]{t1.c0} has a \lstinline[style=codestyle]{NULL} value.
Therefore, in the case of dependent expressions, we need to determine in advance whether to use the \lstinline[style=codestyle]{JOIN} clause in the original query during the generation of $\phi$. This corresponds to line~\ref{alg:phigen} in Algorithm~\ref{alg:oraclealg}. Subsequently, this information will be stored within the set $\{t_i\}$.
Conversely, if the folded expression $\phi$ is used as the predicate of \lstinline[style=codestyle]{JOIN} clause, there is no need for a \lstinline[style=codestyle]{JOIN} clause in the auxiliary query, as the expression $\phi$ would be evaluated with the row values before the \lstinline[style=codestyle]{JOIN} operation.
% If the \lstinline[style=codestyle]{JOIN} were omitted in the auxiliary query, the keys computed would be incorrect, which, in turn, would result in an incorrect oracle.
Second, although we generate non-empty tables, an empty result can still occur, for example, when using an  \lstinline[style=codestyle]{INNER JOIN} with a false predicate. In such scenarios, we discard the test.

\paragraph{Constant propagation.}
To represent the mapping obtained by the auxiliary query, we use a \lstinline[style=codestyle]{CASE} expression in the folded query, as shown in red in step \circled{5} of Figure~\ref{fig:approach}.
This mapping shows that $\phi$ generates \lstinline[style=codestyle]{0} for the first row and \lstinline[style=codestyle]{1} for the second row of the table.
In Listing~\ref{fig:correlated}, the query \circled{F} also uses the \lstinline[style=codestyle]{CASE} expression to represent the mapping for the correlated subquery.
The \lstinline[style=codestyle]{CASE} pattern resembles a well-known optimization for dynamically-typed languages, known as a \emph{polymorphic inline cache}~\cite{holzle1991optimizing}. %, which subsequently enables constant folding and propagation.

\begin{lstlisting}[style=codestyle, belowskip=0pt, caption={The subquery, when used as the fetch keyword in a {\lstinline[style=codestyle]{SELECT}}, must return only one column and one row.}, label={fig:subqueryshape}, float, floatplacement={tb}]
CREATE TABLE t0(c0 INT);
CREATE TABLE t1(c0 INT);
INSERT INTO t0 VALUES (1);
INSERT INTO t0 VALUES (2), (3);
SELECT t0.c0, (SELECT t1.c0 FROM t1 WHERE t1.c0 > t0.c0) FROM t0; 
    -- Error: Subquery returns more than 1 row
SELECT t0.c0, (SELECT t1.c0, t1.c0 FROM t1 WHERE t1.c0=2) FROM t0; 
    -- Error: Operand should contain 1 column(s)
\end{lstlisting}

\subsection{Construction of the Original Query}
\label{sect:queryconstruct}
The expression $\phi$ generated in step \circled{2} can be used in a predicate, which, in turn, can be used in any clause requiring a predicate to construct the original query in step \circled{4}.
In Algorithm~\ref{alg:oraclealg}, the function \texttt{QueryGenerate}, called in line~\ref{alg:orggen}, generates the original query.
%Both independent and dependent expressions can be used in the predicate. However, when dealing with dependent expressions, it's important to ensure that any column from the outer context that is referenced must be explicitly specified in the \lstinline[style=codestyle]{FROM} or JOIN \lstinline[style=codestyle]{clauses}.

\paragraph{Predicate construction.}
%The expressions can be used either on their own or as arguments of functions or operators to construct predicates.
%SQLancer provides a random predicate generator. 
We randomly generate predicates that contain or correspond to $\phi$.  %in abstract syntax tree (AST) format, and then integrate $\phi$ into this AST to construct the predicate.
%Basic expressions and constant expressions can be treated as literal values in predicate construction, since they yield a scalar value for each row in the columns. 
%Literal values can be used in many components, for example, binary operators, and string-related functions.
While we use an existing random generation approach implemented in SQLancer~\cite{rigger2020testing, rigger2020detecting}, subqueries require additional attention, as they were not supported by existing approaches.
Subqueries can evaluate to three different result types: (1) a scalar value, which is a single value; (2) a row value, which is an ordered list of two or more scalar values; (3) multiple row values.
When using a subquery in the fetch clause of a \lstinline[style=codestyle]{SELECT} statement, DBMSs typically allow the subquery only to return a scalar, as illustrated for MySQL in Listing~\ref{fig:subqueryshape}.
This restriction applies to auxiliary queries for dependent expressions generated in step~\circled{3}.
To this end, for correlated subqueries, we either use an aggregate function without using a \lstinline[style=codestyle]{GROUP BY} clause, or use a \lstinline[style=codestyle]{LIMIT} clause. Both ensure that a scalar value is returned.
For both non-correlated and correlated subqueries, we can use one of the multiple subquery operators, which apply for subqueries returning any of the result types.
Such operators include \lstinline[style=codestyle]{EXISTS}, \lstinline[style=codestyle]{IN}, \lstinline[style=codestyle]{ANY}, and \lstinline[style=codestyle]{ALL}.
%The \lstinline[style=codestyle]{EXISTS} operator enables us to test subqueries with empty results and can be used for all outcomes of subqueries and behaves as follows: it equals to a \lstinline[style=codestyle]{FALSE} constant when applied to an empty result subquery and equals to a \lstinline[style=codestyle]{TRUE} constant with other non-empty subqueries.
%The \lstinline[style=codestyle]{IN}, \lstinline[style=codestyle]{ALL}, and \lstinline[style=codestyle]{ANY} operator can be used with non-empty result subqueries, and they accept a subquery or a value list as an argument. 

\paragraph{Query construction.}
The generated predicate can be used in any SQL statement where a predicate is required.
Our approach supports placing these predicates not only in the \lstinline[style=codestyle]{WHERE}, \lstinline[style=codestyle]{JOIN}, \lstinline[style=codestyle]{HAVING}, \lstinline[style=codestyle]{GROUP BY}, and \lstinline[style=codestyle]{ORDER BY} clauses of \lstinline[style=codestyle]{SELECT}, but also in other statements that require predicates, such as \lstinline[style=codestyle]{CREATE INDEX}, \lstinline[style=codestyle]{CREATE VIEW}, \lstinline[style=codestyle]{UPDATE}, \lstinline[style=codestyle]{INSERT}, and \lstinline[style=codestyle]{DELETE}.

\begin{lstlisting}[style=codestyle, belowskip=0pt, caption={A bug found in TiDB caused by a predicate generating an incorrect result in {\lstinline[style=codestyle]{INSERT}}.}, label={fig:insertsubquery}, float, floatplacement={tb}]
CREATE TABLE t0(c0 BIGINT NOT NULL);
INSERT INTO t0(c0) VALUES (1);
CREATE TABLE ot0 (c0 BIGINT);
INSERT INTO ot0 @SELECT t0.c0 AS c0 FROM t0 WHERE VERSION() >= t0.c0@;
|`\circled{O}|` SELECT * FROM ot0; -- empty result |`$\vcenter{\hbox{\includegraphics[height=\baselineskip]{fig/bug.png}}}$|`
|`\circled{A}|` SELECT t0.c0 AS c0 FROM t0 WHERE VERSION() >= t0.c0; -- 1
|`\circled{F}|` SELECT * FROM (@SELECT 1@) AS ft0; -- 1 |`$\vcenter{\hbox{\includegraphics[height=\baselineskip]{fig/greentick.png}}}$|`
\end{lstlisting}
\begin{lstlisting}[style=codestyle, belowskip=0pt, caption={A bug found in CockroachDB caused by a false predicate being always evaluated to true.}, label={fig:casewhenexam}, float, floatplacement={tb}]
CREATE TABLE t1 (v VARBIT);
INSERT INTO t1 VALUES (B'11');
|`\circled{O}|` WITH t2 AS (@SELECT NULL AS b@) SELECT t1.v FROM t1, t2 WHERE t1.v NOT BETWEEN 
    t1.v AND (CASE WHEN NULL THEN t2.b ELSE t1.v END); -- empty result |`$\vcenter{\hbox{\includegraphics[height=\baselineskip]{fig/greentick.png}}}$|`
|`\circled{A}|` SELECT NULL AS b; -- NULL
CREATE TABLE t2 (b BIT);
INSERT INTO t2 VALUES (@NULL@);
|`\circled{F}|` SELECT t1.v FROM t1, t2 WHERE t1.v NOT BETWEEN t1.v AND 
    (CASE WHEN NULL THEN t2.b ELSE t1.v END); -- '11' |`$\vcenter{\hbox{\includegraphics[height=\baselineskip]{fig/bug.png}}}$|`
\end{lstlisting}

\paragraph{Implementation details.}
%To generate semantically correct queries, certain rules must be followed.
Some DBMSs follow strict data type rules when using binary operators, for example, DuckDB and CockroachDB.
Before constructing the predicate, it is necessary to query the return type of the expressions. DuckDB provides the \lstinline[style=codestyle]{typeof()} function for this purpose, whereas CockroachDB provides \lstinline[style=codestyle]{pg_typeof()}.
Additionally, DBMSs such as CockroachDB lack automatic implicit casts for converting predicates from any type to boolean. Therefore, we must generate a boolean expression explicitly as the predicate.
Some DBMSs, such as SQLite and MySQL, can automatically convert the data type of expressions used as operands of operators. This allows us the freedom to choose operators for the expression $\phi$ when testing these DBMSs.

The \lstinline[style=codestyle]{ALL} and \lstinline[style=codestyle]{ANY} operators are not supported in SQLite and DuckDB, and they disallow a value list as an argument in MySQL and TiDB. To overcome this limitation in MySQL and TiDB, we use the \lstinline[style=codestyle]{UNION} operator. For example, we can represent a set of values \lstinline[style=codestyle]{[1, 2, 3]} as \lstinline[style=codestyle]{SELECT 1 UNION SELECT 2 UNION SELECT 3}.
% We also need to ensure that the left operands of these operators have the same number and type corresponding to the right operators.

\subsection{Language Features Beyond Predicates}
\label{sect:extension}
Our approach is applicable not only to constant-folding predicate expressions, but also to contexts that expect a relation (\eg a table or a view reference).
A subquery computing a non-empty result can be used as the source of values for tables, allowing us to test other language features such as \lstinline[style=codestyle]{INSERT}, common table expressions, and derived tables.
In step \circled{4}, we construct the original query by referencing an \textit{original relation}, whose values are obtained from a subquery. In step \circled{5}, we change the reference of the original relation in the original query to a \textit{folded relation} to construct the folded query, whose values are sourced from a table value constructor.  
A table value constructor in SQL is a constant list, typically represented by \lstinline[style=codestyle]{VALUES (...)}. 
Similar to the process of testing predicates, this approach also applies constant folding and constant propagation to subqueries.
As before, the folded query should yield identical results to the original query.

\approach introduces three approaches to constructing original (\ie step \circled{4}) and folded (\ie step \circled{5}) relations. We randomly select one of them to construct the original relation and another one for the folded relation.
The first approach creates a table based on the structure of the subquery results and then uses the subquery in an \lstinline[style=codestyle]{INSERT} statement to add values to the table.
For example, the query \circled{O} shown in Listing~\ref{fig:insertsubquery} references the original relation \lstinline[style=codestyle]{ot0}, whose value originates from the subquery in the \lstinline[style=codestyle]{INSERT} statement.
The second one is a derived table, which is used in the \lstinline[style=codestyle]{FROM} clause of \lstinline[style=codestyle]{SELECT}. 
The query \circled{F} shown in Listing~\ref{fig:insertsubquery} references the folded relation \lstinline[style=codestyle]{ft0}, which is sourced from a constant.
We use a subquery as the operand of \lstinline[style=codestyle]{AS}, as the left operand of \lstinline[style=codestyle]{AS} can only be a subquery.
The third one is common table expression (CTE)~\cite{cebollero2015common}, which was introduced in the SQL Standard of 1999. It is a component of a \lstinline[style=codestyle]{SELECT} query that allows creating a temporary table inside the query and retrieving values from subqueries or table value constructors. 
The query \circled{O}, shown in Listing~\ref{fig:casewhenexam}, demonstrates the application of the CTE in our approach.
The bug-inducing test case shown in Listing~\ref{fig:insertsubquery} uses \lstinline[style=codestyle]{INSERT} statement and derived table,\footnote{\url{https://github.com/pingcap/tidb/issues/\anonymize{43373}}} while the bug-inducing test case shown in Listing~\ref{fig:casewhenexam} uses \lstinline[style=codestyle]{INSERT} statement and CTE.\footnote{\url{https://github.com/cockroachdb/cockroach/issues/\anonymize{102110}}}
% For example, Listing~\ref{fig:insertsubquery} shows a bug found in TiDB triggered by the subquery in \lstinline[style=codestyle]{INSERT} statement.\footnote{https://github.com/pingcap/tidb/issues/\anonymize{43373}} The folded query uses a derived table which enables us to find this bug.
We consider this as an extension of our original approach, because the folded table is not necessarily a constant.

\section{EVALUATION}

We implemented \approach, and evaluated three important aspects. 
First, we evaluated \approach's effectiveness by studying how many unique, unknown bugs our approach could find in widely-used DBMSs that have been extensively tested by the state-of-the-art approaches.
Second, we sought to determine whether existing approaches would indeed be unable to find the bugs we reported.
Third, we compared \approach's performance with the state-of-the-art approaches.

\paragraph{Implementation.}
We implemented \approach in SQLancer, a popular tool for DBMS testing, which supports multiple state-of-the-art test oracles for finding logic bugs, including NoREC, PQS, and TLP. 
SQLancer provides manually-written, rule-based generators specific to the DBMS under test, which can be used to generate statements and expressions.
\approach initializes the database states using SQLancer's random generation method, which generates random statements to create tables, views, indexes, and insert values into tables.
%\approach was easy to implement.
For generating the original query, we  adopted the logic used in NoREC, and added support for generating subqueries.
For generating auxiliary queries, we implemented an additional generator.
For folded queries, we  replaced the expression in the original query with the constant by replacing child nodes in the Abstract Syntax Tree (AST) that SQLancer provides.
%We implemented the generator for auxiliary queries, and adopted the random query generation method used in NoREC to generate the original queries, and this query generation method can also be used to generate subqueries, which are not supported by NoREC.
% For each test case, we start by generating the expression $\phi$ that will undergo constant folding using SQLancer's expression generator. 
% Then, we construct the auxiliary query to obtain its results. 
% Similar to NoREC, we generate the \lstinline[style=codestyle]{SELECT} query by building each clause of it one-by-one. 
% When generating the subquery, all of the predicate are generated by the expression generator of SQLancer. 
% When generating the predicate in original query, we decide whether to embed the expression $\phi$ based on a predefined probability, but also ensure that at least one clause will use $\phi$.
% The expression $\phi$ can be used as a single predicate, or a sub-expression of the predicate.
\approach is easy to understand and implement, so we could have chosen alternative database and query generators that we could similarly extend to realize our approach.

\paragraph{Tested DBMSs.} We selected five DBMSs as our test targets: SQLite,\footnote{\url{https://sqlite.org/index.html}} MySQL,\footnote{\url{https://www.mysql.com/}} CockroachDB~\cite{DBLP:conf/sigmod/TaftSMVLGNWBPBR20}, DuckDB~\cite{raasveldt2020data}, and TiDB~\cite{DBLP:journals/pvldb/HuangLCFMXSTZHW20}.
We chose these specific DBMSs for several reasons. Firstly,  these DBMSs are widely used, popular, and considered mature. 
MySQL and SQLite are long-established DBMSs and thus rank highly in rankings such as the DB-Engines ranking.\footnote{\url{https://db-engines.com/en/ranking}} 
CockroachDB, DuckDB, and TiDB are relatively recent DBMSs, which are highly popular on GitHub, with star counts of 27.2K, 10.5K, and 34.2K respectively.
Secondly, these DBMSs represent various types of DBMSs. 
SQLite and DuckDB are embedded DBMSs that run in the same process as the application that uses them.
MySQL is a traditional, relational, and client-server DBMS. 
Both CockroachDB and TiDB are distributed relational DBMSs, designed to handle large-scale deployments and ensure high availability and scalability in a distributed environment~\cite{DBLP:journals/sigmod/PavloA16}.
Finally, all five DBMSs have been extensively tested by various methods such as NoREC~\cite{rigger2020detecting}, TLP~\cite{rigger2020finding}, PQS~\cite{rigger2020testing}, and DQE~\cite{song2023testing}, suggesting that any newly found bugs might have been overlooked by these existing approaches.

\paragraph{DBMSs versions.}
We tested the latest development versions of the aforementioned DBMSs.
We downloaded SQLite from its source code repository, and downloaded the other DBMSs from GitHub. For SQLite, we tested version \texttt{c1f2a1d5} and later trunk versions. For MySQL, we tested commit version \texttt{ea7087d}. For CockroachDB, we tested commit version \texttt{07c7d4b} and later versions. For DuckDB, we tested commit version \texttt{b8cf6a9} and later versions. For TiDB, we tested version \texttt{c233969} and later commit versions.

\paragraph{Baselines.}
We selected NoREC~\cite{rigger2020detecting}, TLP~\cite{rigger2020finding}, DQE~\cite{song2023testing}, and EET~\cite{DBLP:conf/osdi/Jiang024} as baselines for evaluation.
NoREC, TLP, PQS~\cite{rigger2020testing}, TQS~\cite{DBLP:journals/pacmmod/TangWZ0023}, DQE, and EET are the state-of-the-art approaches for automatically testing DBMSs. 
NoREC and DQE assume that the same predicate always accesses the same row of a table, regardless of which clause it is used in. 
TLP decomposes a query into three partitioning queries, which retrieve rows based on predicates \lstinline[style=codestyle]{p}, \lstinline[style=codestyle]{NOT p}, and \lstinline[style=codestyle]{p IS NULL}, respectively. 
EET is a concurrent work with \approach, which has proposed the first expression-level manipulation approach, known as Equivalent Expression Transformation (EET), which introduces tautologies and contradictions
% determined boolean expressions---constructed using logical operations that always evaluate to either true or false---
to construct equivalent queries.
PQS generates a \lstinline[style=codestyle]{SELECT} query to retrieve a pivot row, and checks whether the DBMS fetches it as expected.
PQS requires a high implementation effort, as operations of the DBMS must be implemented also in the testing tool. For this reason, it is no longer actively maintained in the SQLancer project, and currently triggers false alarms.\footnote{\url{https://github.com/sqlancer/sqlancer/issues/527}} %Other approaches also have not been compared against it.
TQS was proposed to find bugs in join optimizations; however, it is not publicly available.
Thus, we omitted both PQS and TQS from the comparison.
% Therefore, we selected NoREC, TLP, DQE, and EET as baselines for evaluation.

\subsection{Effectiveness}
\paragraph{Methodology.}
We evaluated the effectiveness of \approach by testing the five aforementioned DBMSs.
We intermittently ran \approach for a period of four months, during which we also implemented it. 
This is a standard methodology used to evaluate the effectiveness of automated testing tools~\cite{rigger2020detecting, DBLP:conf/oopsla/LeSS15, DBLP:conf/sigsoft/MansurCW21}.
Before reporting bugs, we manually reduced the bug-inducing test cases~\cite{DBLP:journals/tse/ZellerH02}.
To avoid reporting duplicate issues, for multiple reports that we suspected to affect the same component, we reported the subsequent bug only after the previously reported one was fixed. 
For some DBMSs (\eg TiDB), we refrained from reporting more bugs, due to the large number of unfixed bugs.

\begin{table*}[tb]\small
    \centering
    \caption{\approach found \totalbug unique bugs in five mature DBMSs.}
    \label{tab:totalbugstatis}
    \begin{tabular}{l r r r r r r r} \toprule
         \multirow{2}{*}{DBMS}& \multicolumn{4}{c}{Bug type} & & \multicolumn{2}{c}{Bug status}\\\cline{2-5}\cline{7-8}
         & Logic bug & Internal error & Crash & Hang & & Fixed & Verified \\\midrule
         SQLite & 6 & 1 & 0 & 0 & & 7 & 0 \\
         MySQL & 1 & 1 & 0 & 0 & & 0 & 2 \\
         CockroachDB & 7 & 4 & 0 & 2 & & 11 & 2 \\
         DuckDB & 5 & 2 & 2 & 3 & & 12 & 0 \\
         TiDB & 5 & 6 & 0 & 0 & & 3 & 8 \\
         \midrule
         Total & 24 & 14 & 2 & 5 & & 33 & 12 \\\bottomrule
    \end{tabular}
\end{table*}

% \begin{table}[tb]\small
%     \centering
%     \caption{\approach found \totalbug unique bugs in five mature DBMSs.}
%     \label{tab:totalbugstatis}
%     \begin{tabular}{l r r r r} \toprule
%          \multirow{2}{*}{DBMS}& \multicolumn{4}{c}{Bug type}\\\cline{2-5}
%          & Logic bug & Internal error & Crash & Hang \\\midrule
%          SQLite & 6 & 1 & 0 & 0\\
%          MySQL & 1 & 1 & 0 & 0\\
%          CockroachDB & 7 & 4 & 0 & 2\\
%          DuckDB & 5 & 2 & 2 & 3\\
%          TiDB & 5 & 6 & 0 & 0\\
%          \midrule
%          Total & 24 & 14 & 2 & 5\\\bottomrule
%     \end{tabular}
% \end{table}

% \begin{table}[tb]\small
%     \centering
%     \caption{The status of bugs found by \approach.}
%     \label{tab:bugstatis}
%     \begin{tabular}{l r r r} \toprule
%          DBMS & Fixed & Verified & Open \\\midrule
%          SQLite & 7 & 0 & 0 \\
%          MySQL & 0 & 2 & 0 \\
%          CockroachDB & 9 & 3 & 1 \\
%          DuckDB & 9 & 2 & 1\\
%          TiDB & 2 & 9 & 0\\
%          \midrule
%          Total & 27 & 16 & 2 \\\bottomrule
%     \end{tabular}
% \end{table}

\paragraph{Results.}
Table~\ref{tab:totalbugstatis} shows the number of bugs found by \approach, as well as the status of the bugs. 
\approach found a total of \totalbug previously unknown bugs. Out of these, \totallogicbug were logic bugs, 14 were internal errors, 2 were crashes, and 5 were hang-related issues. 
% Table~\ref{tab:bugstatis} shows the statuses of all bugs. 
Out of all \totalbug bugs, \fixedbug were fixed, and \conbug were verified.
% , and \openbug reports are waiting for confirmation.
%\approach found \totallogicbug previously unknown unique logic bugs. 
Out of the \totallogicbug logic bugs, \fixedlogicbug were fixed, and \conlogicbug were verified.
% , and \openlogicbug reports are waiting for confirmation. 
These results are highly encouraging, considering that these DBMSs have been the focus of many testing works as mentioned above.

\paragraph{Logic bugs causes}
% Table~\ref{tab:logicbugstatis} presents the logic bug statuses. 
We analyzed the queries that triggered the bugs and identified that 12 bugs were triggered by folded queries---the queries generated and executed in step \circled{5}--- and 12 by original queries. 
Out of the bug-inducing folded queries, 11 queries used folded constants that were derived from non-correlated subqueries.
Of the bugs caused by them, 6 were triggered by replacing a constant, and the remaining 5 bugs were triggered by a value list, all of which were related to the \lstinline[style=codestyle]{IN} operator.
Only one query used a constant that was derived from the query with a simple expression.
Out of the remaining 12 bugs triggered by original queries, all of them used queries containing non-correlated subqueries. 
Feedback from the developers indicated that 3 bugs were bugs in subquery processing, while 5 bugs were unrelated to subqueries. For the 4 remaining bugs, we are unclear about the root cause due to limited developer feedback.
Overall, we detected most of the bugs through non-correlated subqueries or folded queries derived from non-correlated subqueries. 
This is primarily because non-correlated subqueries were our initial test focus, as they can be executed independently and are straightforward to implement. 
Additionally, half of all the bugs were triggered by the folded queries. We believe that applying \approach to basic expressions or constant expressions could also have been used to detect some of these bugs.
For example, for the bug-inducing test case shown in Listing~\ref{fig:sqlitejoin}, using a constant expression \lstinline[style=codestyle]{1 - 1} as $\phi$ could have detected this bug.

\paragraph{Other bugs.}
\approach found a total of \totalotherbug other bugs in five DBMSs, including crashes, internal errors, and hangs. Out of the \totalotherbug bugs, \fixedotherbug have been fixed, and \conotherbug have been verified. 
We found two crashes in DuckDB, both of which caused our tool to terminate unexpectedly, and the test cases resulted in segmentation faults when using DuckDB's command line interface. These two crashes were introduced in the IEJoin optimization~\cite{DBLP:journals/pvldb/KhayyatLSOPQ0K15}. One crash was caused by an index out-of-bounds error, while the other was due to a type mismatch. These two crashes have been fixed.
% \begin{figure}[t]
%     \centering
%     \captionsetup{skip=4pt}
%     \parbox{\linewidth}{
% 	\lstset{style=codestyle}
% 	\lstinputlisting[morekeywords={WITH},basicstyle=\scriptsize\ttfamily\bfseries]{code/tidberrors.sql}
%     }
%     \caption{Two internal errors found in TiDB.}
%     \label{fig:tidberror}
% \end{figure}
% Figure~\ref{fig:tidberror} illustrates two internal errors we found in TiDB.\footnote{https://github.com/pingcap/tidb/issues/\anonymize{44673}, https://github.com/pingcap/tidb/issues/\anonymize{44689}} 
% The first error was triggered by a common table expression, which we discovered when applying \approach to a subquery. 
% The second error was caused by the \lstinline[style=codestyle]{CHECK} clause, which could also be triggered in MySQL. 
% To minimize the manual effort required for checking internal error bug reports, we maintain an error message list to track and record the reported errors, and ignore the duplicate reports.
We found 14 internal errors in these five DBMSs, 6 have been fixed, and the others were verified.
All 5 hang-related bugs found in DuckDB and CockroachDB have been fixed.
We believe these crashes, internal errors, and hangs could  be detected by other automated testing approaches~\cite{DBLP:conf/ccs/ZhongC0ZLW20, DBLP:conf/kbse/FuLWWJ22}.

% \begin{table}[htbp]\small
%     \centering
%     \caption{\approach found \totalotherbug unique other bugs in four DBMSs.}
%     \label{tab:otherbugs}
%     \begin{tabular}{l r r r} \toprule
%          DBMS & Crash & Internal Error & Hang \\\midrule
%          SQLite & 0 & 1 & 0 \\
%          CockroachDB & 0 & 4 & 1 \\
%          DuckDB & 2 & 2 & 3 \\
%          TiDB & 1 & 3 & 0\\
%          \midrule
%          Total & 3 & 10 & 4 \\\bottomrule
%     \end{tabular}
% \end{table}

\paragraph{Bug importance.}
Although anecdotal, developer feedback is an important indicator of an approach's effectiveness and the found bugs' importance.
Two DBMS companies reached out to us about our testing efforts; both were interested in how we found the bugs, and one invited us to present the approach to the development team.
Furthermore, we received positive feedback on the public bug trackers.
For example, a developer of CockroachDB commented in one of our reports ``\emph{we really appreciate your work!}''\footnote{\url{https://github.com/cockroachdb/cockroach/issues/\anonymize{104319}}}
Three bugs in CockroackDB were assigned the ``S-0'' or ``S-2'' label, indicating high-impact bugs that were difficult to resolve.
Three bug reports in TiDB were labeled as ``Major'', which represents the highest bug severity.
% Six of our reports were verified as bugs with many users impacted and difficult to fix.
%Our bug reports significantly aided DBMS developers in uncovering numerous critical and elusive bugs. These reports have garnered substantial developer interest. TiDB and CockroachDB developers expressed keen curiosity about \approach and expressed anticipation for delving into it.

\paragraph{False alarms} 
While realizing our approach, we identified corner cases that resulted in false alarms in our initial implementation. First, applying constant folding with floating-point numbers can result in false alarms. We avoid these in practice by eschewing test cases with small or large float-point values. Second, SQLite's relaxed type system allows, in some context, values of different types to be returned. We avoid this through explicit casts to the data type we expect.
We have not observed any false alarms after addressing these issues.

\paragraph{Discussion.}
Although we intermittently ran \approach over a period of four months, during which we also implemented it, most of the bugs were found at the beginning of \approach's execution. As shown in the evaluation in Section~\ref{sect:comparison}, CODDTest identified 25 unique bugs in an older version of SQLite within 24 hours, with 13 of those bugs found in the first hour. Since \approach is a black-box method, it is an ongoing research challenge to determine whether a system has been sufficiently tested~\cite{DBLP:conf/icse/LiyanageBTL23}.
% challenging to determine if a DBMS has been sufficiently tested, necessitating the continuous running of \approach.

\subsection{Test Oracle Comparison}
\label{sect:comparison}
A testing approach is valuable if it finds bugs that were overlooked by existing approaches. Thus, we sought to confirm whether the state-of-the-art test oracles are indeed ineffective in finding the bugs found by \approach,
as well as to identify the types of bugs that \approach can not detect.

\begin{table}[bt]\small
    \centering
    \caption{The number of detectable bugs by test oracles.}
    \label{tab:compstat}
    \begin{tabular}{l c c c c} \toprule
         Oracles & NoREC & TLP & DQE & Only by \approach \\\midrule
         Num & 11 & 12 & 4 & 11 \\\bottomrule
    \end{tabular}
    % \vspace*{-5pt}
\end{table}

\paragraph{Methodology.}
% To address these challenges, 
To address these two questions, we designed three experiments. First, 
we ran \approach on the release version 3.30.0 of SQLite, in which NoREC and TLP detected 31 bugs, to determine how many of these bugs \approach could identify over 24 hours with 10 threads.
The developers of SQLite were proactive in fixing bugs, which allowed us to determine whether two bug reports trigger the same bug by applying the bug-fixed commit to SQLite and checking if the bug could still be triggered. 
We selected NoREC, TLP, and EET for this comparison, because DQE is conceptually similar to NoREC. Additionally, the test cases generated by DQE are incomplete and cannot be automatically analyzed.
Second, 
we analyzed the earliest versions of DBMSs where logic bugs identified by \approach can be triggered, to determine whether these bugs were introduced before the state-of-the-art approaches were published, suggesting that the existing approaches missed the bugs that \approach subsequently found.
This methodology was also used in the EET work~\cite{DBLP:conf/osdi/Jiang024}.
Third,
we implemented a best-effort comparison by manually inspecting and analyzing the bug-inducing test cases and bugs found by \approach and analyzing whether the state-of-the-art test oracles could have found them. 
We selected NoREC, TLP, and DQE for this comparison, because EET's transformations explore an extensive search space, making it difficult to conduct the transformations and check their results manually.
%We believe this approach is reasonable, and it was applied to the evaluation of prior works such as TLP~\cite{rigger2020finding} and NoREC~\cite{rigger2020detecting}.
For every analyzed bug, we include an analysis in the supplementary materials including the test cases constructed by the test oracles, allowing scrutinization of the results.

\paragraph{Results of oracles on an older version of SQLite}
During a 24-hour period, NoREC, TLP, EET and \approach generated 97,003, 103,103, 16, and 6,990 bug reports, respectively, with 27, 27, 6, and 25 of the bugs being unique. Additionally, NoREC, TLP, EET, and \approach reported 3, 2, 3, and 4 bugs that were found by that oracle alone.
Although the significant overlap might be surprising, we believe it is reasonable---a testing approach is useful if it can find new bugs overlooked by other approaches.
Note that we conducted this experiment on a single and stable version of SQLite and that in other, especially older versions of SQLite, likely more unique bugs could be detected.
% \approach has generated 6,990 logic bug reports, resulting in 25 unique bugs after deduplication.
% Among these 25 unique bugs, 13 were reported by NoREC, 1 by TLP, 2 by other logic bug oracles, and 9 were identified through crashes and assertion failures.

As all of these three oracles are black-box testing methods, it is challenging to determine which types of bugs will be overlooked in general.
However, we noticed that \approach overlooked 14 bugs, 4 of which were related to indexing. Therefore, we speculate that \approach cannot effectively test indexing functionality.

% Additionally, it would be unfair to conclude that \approach cannot detect a certain bug if it does not find it within 24 hours. 
% We found that the majority of bug reports generated by these three oracles were repeatedly reporting subsets of bugs, with many bugs having fewer than 10 reports. Specifically, 
Although it is an established methodology to run testing approaches for 24 hours~\cite{10.1145/3243734.3243804}, we noticed that it might be insufficient for some bugs to be consistently found. By running the 24-hour experiments, we found that
9, 13, and 16 bugs found by NoREC, TLP, and \approach, respectively, had fewer than 10 reports. Therefore, generating certain corner cases might require a substantial amount of time. This motivated us to conduct the manual comparison study, whose results are detailed next.

\paragraph{Results on bugs introduction times}
\approach can detect long-latent bugs in DBMSs that have been overlooked by state-of-the-art approaches.
NoREC supports SQLite and MySQL. \approach found 7 logic bugs in these two systems, and 1 bug was introduced before NoREC was published. 
All DBMSs supported by \approach are also supported by TLP; 6 out of the 24 bugs found by \approach were introduced before TLP was published. 
DQE supports SQLite, MySQL, CockroachDB, and TiDB, and 15 out of 19 bugs found in these DBMSs were introduced before DQE was published. 
EET supports MySQL, SQLite, and TiDB, and the 12 bugs found in these DBMSs were all introduced prior to EET's first bug report, indicating that EET may have missed these bugs.
Out of the 24 logic bugs found by \approach, 6 were introduced before 2020, and 20 were introduced before 2023. The bug with the longest latency---14 years---was found in MySQL. This result demonstrates that \approach can effectively identify long-latent bugs.

\paragraph{Results of manual comparison.}
As shown in the manual-analysis results in Table~\ref{tab:compstat}, out of \totallogicbug bugs detected by \approach, 11 can be detected by NoREC, 12 can be detected by TLP, 4 can be detected by DQE, and 11 can only be detected by \approach.
% We found 7 bugs in language features not covered by existing approaches, including subqueries, \lstinline[style=codestyle]{ON},   \lstinline[style=codestyle]{CASE}, and \lstinline[style=codestyle]{INSERT} statements.
We found that these 11 bugs were in language features that were not supported by existing test oracles, including subqueries, \lstinline[style=codestyle]{ON} clauses, \lstinline[style=codestyle]{CASE} and \lstinline[style=codestyle]{ANY} expressions, the \lstinline[style=codestyle]{AVG} function, and \lstinline[style=codestyle]{INSERT} statements.

\paragraph{Subqueries.}
Three bugs related to subqueries, which are out-of-scope for the three test oracles.
Listing~\ref{fig:illustexample} shows such a bug-inducing test case for SQLite. 
The SQLite developers communicated that reproducing it requires five conditions: 
(1) the query must contain an aggregate subquery; 
(2) the aggregate subquery must have a \lstinline[style=codestyle]{GROUP BY} clause; 
(3) the \lstinline[style=codestyle]{GROUP BY} clause must reference terms that are not included in the result set of the query;
(4) the query planner must choose to implement the \lstinline[style=codestyle]{GROUP BY} clause by doing a sort operation;
(5) the outer query that contains the aggregate subquery must make use of indexed expressions.
Under these conditions, SQLite assigned an incorrect value to a variable in the SQL AST associated with the subquery.
The transformations performed by existing test oracles are incapable of generating queries that violate these conditions, preventing them from detecting this bug.
We found the second subquery-related bug in DuckDB, which was caused by incorrect processing of the return type of a subquery. \approach executed the auxiliary query to obtain the subquery's results with the correct data type, which were then used in the folded query. The folded query subsequently generated the correct results. However, due to the incorrect handling of the subquery's return type, the original query produced incorrect results. 
The third one was found in TiDB and caused by the incorrect recognition of columns with identical names, resulting in the misinterpretation of a non-correlated subquery as a correlated one.

\begin{lstlisting}[style=codestyle, belowskip=0pt, caption={A bug found in SQLite related to {\lstinline[style=codestyle]{JOIN}}.}, label={fig:sqlitejoin}, float, floatplacement={tb}]
CREATE TABLE vt0(c2);
CREATE TABLE t1(c0 TEXT);
INSERT INTO t1(c0) VALUES (1);
INSERT INTO vt0(c2) VALUES (-1);
CREATE VIEW v0(c0) AS SELECT 0 FROM t1;
|`\circled{O}|` SELECT vt0.c2 AS c1 FROM t1 CROSS JOIN v0 ON ( 
    @EXISTS (SELECT v0.c0 FROM v0 WHERE false)@) FULL OUTER JOIN vt0 ON 1; -- -1 |`$\vcenter{\hbox{\includegraphics[height=\baselineskip]{fig/greentick.png}}}$|`
|`\circled{A}|` SELECT v0.c0 FROM v0 WHERE false;-- empty result
|`\circled{F}|` SELECT vt0.c2 AS c1 FROM t1 CROSS JOIN v0 ON (@0@) FULL OUTER JOIN vt0 ON 1; 
    -- empty result |`$\vcenter{\hbox{\includegraphics[height=\baselineskip]{fig/bug.png}}}$|`
\end{lstlisting}

\paragraph{\lstinline[style=codestyle]{ON} clause.}
The \lstinline[style=codestyle]{ON} clause of \lstinline[style=codestyle]{JOIN} remained untested by NoREC, TLP, and DQE. Since these three test oracles rely on the direct mapping relationship between predicates and corresponding rows, which cannot be directly determined for the \lstinline[style=codestyle]{JOIN ON} clause, they are unable to effectively test the clause. 
%This is because the predicate within the \lstinline[style=codestyle]{JOIN ON} clause does not have a direct mapping to the results.
Two bugs in SQLite found by \approach were missed by these three test oracles for this reason.
Listing~\ref{fig:sqlitejoin} shows one of the bugs that was triggered when replacing a predicate in the \lstinline[style=codestyle]{JOIN ON} clause with a constant.\footnote{\url{https://sqlite.org/forum/forumpost/\anonymize{96cd4a7e9e}}} The original query that embedded the subquery generated the correct results and provided an opportunity to find this bug.

\paragraph{\lstinline[style=codestyle]{CASE} and \lstinline[style=codestyle]{ANY} expressions.}
Listing~\ref{fig:casewhenexam} shows a bug found in CockroachDB, where a bug in processing the \lstinline[style=codestyle]{CASE} caused the expression to be incorrectly evaluated to \lstinline[style=codestyle]{TRUE}, regardless of in which clause the predicate was placed. Consequently, NoREC, TLP, and DQE were unable to detect this bug.
We observed two bugs related to \lstinline[style=codestyle]{CASE} expressions and one bug related to \lstinline[style=codestyle]{ANY} expressions, which consistently evaluated to an incorrect result.

\paragraph{\lstinline[style=codestyle]{AVG} function.}
A bug in CockroachDB related to \lstinline[style=codestyle]{AVG} function, which produced inconsistent results when the argument order was altered.
The original and folded queries produced results with different orders, enabling us to find this bug.

\paragraph{Language features beyond queries.}
Listing~\ref{fig:insertsubquery} shows a bug found in TiDB, specifically in an \lstinline[style=codestyle]{INSERT} statement, whose argument was a non-correlated subquery. The inner query may return a non-empty result, but it fails to produce a non-empty result within the \lstinline[style=codestyle]{INSERT} statement. None of the existing test oracles support testing \lstinline[style=codestyle]{INSERT} statements, as \lstinline[style=codestyle]{INSERT} does not directly use predicates.
% We attempted to use the predicate from the subquery in a \lstinline[style=codestyle]{DELETE} statement, such as \lstinline[style=codestyle]{DELETE FROM t0 WHERE VERSION()>=t0.c0;}. Interestingly, DQE indicates a bug when using the predicate in a  \lstinline[style=codestyle]{DELETE} clause, and we speculate that the root cause is the same in both cases.
The extension of our method described in Section~\ref{sect:extension}, can find bugs in \lstinline[style=codestyle]{INSERT} statements even when the bug cannot be reproduced using other statements.

\begin{lstlisting}[style=codestyle, caption={A bug found in CockroachDB related to {\lstinline[style=codestyle]{IN}} operator.}, label={fig:cockroachinexample}, float, floatplacement={tb}]
CREATE TABLE t (c INT4);
INSERT INTO t (c) VALUES (0);
|`\circled{F}|` SELECT c FROM t WHERE c IN (0, 862827606027206657::INT8); -- empty result |`$\vcenter{\hbox{\includegraphics[height=\baselineskip]{fig/bug.png}}}$|`
\end{lstlisting}

\begin{lstlisting}[style=codestyle, caption={A bug found in TiDB related to {\lstinline[style=codestyle]{IN}} operator.}, label={fig:tidbinexample}, float, floatplacement={t}, belowskip=-0.6 \baselineskip]
CREATE TABLE t0(c0 BOOL);
INSERT INTO t0 VALUES (true);
|`\circled{F}|` SELECT t0.c0 FROM t0 WHERE ((CASE ((CASE t0.c0 WHEN 6 THEN 0.03 ELSE t0.c0 END ) 
    LIKE (t0.c0)) WHEN t0.c0 THEN 1 END) AND (t0.c0 IN (1))); -- empty result |`$\vcenter{\hbox{\includegraphics[height=\baselineskip]{fig/bug.png}}}$|`
\end{lstlisting}

\paragraph{Variations in expression behavior across clauses.}
NoREC and DQE were designed to identify logic bugs by assuming that a specific predicate evaluates to the same value irrespective of in which clause it is used. However, we discovered that either DQE or NoREC failed to detect some of these bugs. 
There are two bugs in this category.
Listing~\ref{fig:cockroachinexample} illustrates a bug in CockroachDB associated with the \lstinline[style=codestyle]{IN} operator.\footnote{\url{https://github.com/cockroachdb/cockroach/issues/\anonymize{102864}}} CockroachDB produced the correct results when the right operand was a subquery, but yielded incorrect results when the right operand was a value list. 
This predicate produced incorrect results in \lstinline[style=codestyle]{SELECT} queries, while correctly functioning in \lstinline[style=codestyle]{UPDATE} and \lstinline[style=codestyle]{DELETE} statements. As a result, NoREC failed to detect this bug, whereas DQE successfully found it.
Listing~\ref{fig:tidbinexample} illustrates a bug in TiDB that is also associated with the \lstinline[style=codestyle]{IN} operator.\footnote{\url{https://github.com/pingcap/tidb/issues/\anonymize{43624}}} Similar to the bug mentioned above, it generated correct results when the right operand was a subquery, but yielded incorrect results when the right operand was a value list. 
The difference, however, lies in the behavior exhibited in \lstinline[style=codestyle]{WHERE} clauses, where it consistently produced incorrect results, while correctly functioning when used in the fetch clause of a \lstinline[style=codestyle]{SELECT}. As a result, DQE failed to detect this bug, while NoREC successfully found it.

\begin{lstlisting}[style=codestyle, caption={A bug found in DuckDB, which triggers an error when applying NoREC to it. The expressions highlighted in blue illustrate the application of the NoREC oracle.}, label={fig:duckdberror}, float, floatplacement={t}, belowskip=-0.6 \baselineskip]
CREATE TABLE t0(c1 INT8);
INSERT INTO t0(c1) VALUES ((1));
|`\circled{O}|` SELECT t0.c1 FROM t0 WHERE ^(((-1314689763) + (-1947665992)) <=^ 
    ^(EXISTS( SELECT t0.c1 FROM t0 WHERE false)))^; -- empty result |`$\vcenter{\hbox{\includegraphics[height=\baselineskip]{fig/bug.png}}}$|`
|`\circled{A}|` SELECT t0.c1 FROM t0 WHERE false;-- empty result
|`\circled{F}|` SELECT t0.c1 FROM t0 WHERE (((-1314689763) + (-1947665992)) <= (false)); -- 1 |`$\vcenter{\hbox{\includegraphics[height=\baselineskip]{fig/greentick.png}}}$|`
SELECT ^((-1314689763)+(-1947665992)) <= (EXISTS( SELECT t0.c1 FROM t0 WHERE false))^ 
    FROM t0;
    -- Out of Range Error: Overflow in addition of 
    -- INT32 (-1314689763 + -1947665992)!
\end{lstlisting}

Furthermore, four bug-inducing test cases triggered errors when other oracles were applied to them, indicating that these logic bugs could not be found by them. 
Two of them triggered internal errors when we applied NoREC to them, and two of them triggered semantic errors when we applied DQE to them.
We encountered scenarios where a predicate triggered a logic bug within \lstinline[style=codestyle]{WHERE} clauses, while simultaneously causing an internal error when used in a \lstinline[style=codestyle]{SELECT} statement.
Listing~\ref{fig:duckdberror} shows one of the two such bugs found by \approach.\footnote{\url{https://github.com/duckdb/duckdb/issues/\anonymize{7094}}}
When we used the predicate triggering the logic bug in the \lstinline[style=codestyle]{SELECT}'s fetch clause, the query triggered an internal error. As a result, although NoREC cannot directly detect this logic bug, the internal error could have still been found.
However, when applying DQE to the logic bug we discovered in TiDB, we observed that the predicate, which executed normally in \lstinline[style=codestyle]{SELECT}, triggered a semantic error when used in \lstinline[style=codestyle]{UPDATE} and \lstinline[style=codestyle]{DELETE} statements.
Although the authors of DQE referenced a similar case found in MySQL in their paper, they did not provide an explanation for the underlying reason. To investigate further, we searched the MySQL bug list and uncovered the root cause.\footnote{\url{https://bugs.mysql.com/bug.php?id=111483}}
In MySQL, the \lstinline[style=codestyle]{SELECT} statement allows comparing values with different data types. However, it does not permit this in \lstinline[style=codestyle]{UPDATE} and \lstinline[style=codestyle]{DELETE} statements.
Therefore, DQE triggers a semantic error in this particular test case, and is unable to detect this logic bug.

% \lstinputlisting[style=codestyle, caption={A bug found in SQLite, related to the JSON function, requires the use of a {\lstinline[style=codestyle]{VIEW}} to be triggered.}, label={fig:sqliteview}, float, floatplacement={t}]{code/sqliteview.sql}

% \paragraph{Limitations of DQE}
% During manual analysis of the bugs and test cases, we observed that DQE could not be applied to the majority of the test cases. 
% Firstly, we encountered four bugs when querying rows from a \lstinline[style=codestyle]{VIEW}. In most DBMSs, performing \lstinline[style=codestyle]{UPDATE} and \lstinline[style=codestyle]{DELETE} operations on a \lstinline[style=codestyle]{VIEW} is not allowed.\footnote{https://www.sqlite.org/lang\_createview.html} 
% Although MySQL does support updatable \lstinline[style=codestyle]{VIEW}s, the numerous constraints associated with it limit DQE's applicability.\footnote{https://dev.mysql.com/doc/refman/8.0/en/view-updatability.html}
% Secondly, we encountered three bugs related to tables created by common table expressions. These tables are valid only within the current statement, making it impossible to perform \lstinline[style=codestyle]{UPDATE} and \lstinline[style=codestyle]{DELETE} operations on them.
% Thirdly, DQE does not support all queries involving \lstinline[style=codestyle]{JOIN}.
% The presence of a \lstinline[style=codestyle]{JOIN} introduces additional constraints on the \lstinline[style=codestyle]{SELECT} operation. To be equivalent to the predicates used in \lstinline[style=codestyle]{UPDATE} and \lstinline[style=codestyle]{DELETE}, the \lstinline[style=codestyle]{SELECT} operation must have the predicate in \lstinline[style=codestyle]{WHERE} clause as the only constraint.
% Fourthly, both MySQL and TiDB do not allow performing \lstinline[style=codestyle]{UPDATE} and \lstinline[style=codestyle]{DELETE} operations on a table with a subquery that selects from the same table.\footnote{https://dev.mysql.com/doc/refman/8.0/en/update.html} 

\paragraph{Summary.}
\approach can find bugs that are missed by the state-of-the-art approaches, such as in subqueries, \lstinline[style=codestyle]{JOIN}, operators, and functions. 
We believe that the existing state-of-the-art test oracles %(\ie NoREC, DQE, and TLP) 
cannot be extended to find these missed bugs in any obvious way. While query generators could still generate the same queries, the existing test oracles would miss the logic bugs associated with these language features.
Additionally, \approach can test language features beyond predicates, as discussed in Section~\ref{sect:extension}. This includes common table expressions (CTE) and derived tables.

\subsection{Efficiency Comparison}
In this section, we compare the performance of \approach to the state-of-the-art approaches.
\paragraph{Methodology.}
NoREC, TLP, and DQE are all implemented based on SQLancer, which enables a fair comparison of the approach, as we could use the same settings (\eg the same number of threads).
Besides \approach, we also evaluated two other configurations to explore the impact of testing different expression types on performance. 
Specifically, we explored the performance of \approach when using only expressions with no subqueries (\ie \approach \& Expression) and only subqueries (\ie \approach \& Subquery).

\paragraph{Metrics.}
We define six metrics for this evaluation.
We measured test throughput by counting the number of successful test cases executed.
We recorded the number of queries that were successfully executed (\ie successful queries) and those that encountered expected errors (\ie unsuccessful queries).
Expected errors either refer to queries triggering unfixed internal errors in the DBMS, or cases where the query is semantically incorrect (\eg errors like unexpected integer overflows are difficult to avoid during construction).
Oracles differ in how many queries they execute for each test case.
Note that this number is not static as, for example, generating a test might fail due to an unsuccessful query.
Thus, we compute the average number of queries for each successfully-executed test, that is, the queries per test (QPT).
We also evaluated the number of unique query plans generated by each method.
% , because recent research suggested that exercising more unique query plans corresponds to exercising more interesting, potentially erroneous behaviors of the DBMS under test~\cite{DBLP:conf/icse/BaR23}.
We collected only the query plan of the most complex query, which was the optimized query in NoREC, the partitioning query in TLP, the \lstinline[style=codestyle]{SELECT} query in DQE, and the original query in \approach.
Lastly, we compare the branch coverage of each oracle at the end of execution, as this provides a more rigorous assessment than statement coverage.\footnote{\url{https://www.sqlite.org/testing.html}}

\paragraph{Experimental setup.}
We conducted the performance evaluation on a server with a 64-Core AMD EPYC 7763 Processor at 2.45GHz and 512GB of memory running Ubuntu 22.04.
We selected SQLite as our test target, and conducted this experiment with release version \texttt{3.42.0.0}.
We executed each approach with 10 threads for a duration of 24 hours and recorded the number of tests conducted.
SQLancer provides a \textit{MaxDepth} option, which controls the maximum depth of an expression.
We used the default configuration of SQLancer, where this option is set to 3.
We conducted separate experiments with the same configuration and times to collect statistics on unique query plans and branch coverage, as querying these metrics requires additional time.

\paragraph{Results.}
\begin{table*}[tb]\small
    \centering
    \caption{The number of tests conducted by each approach.}
    \vspace*{-3pt}
    \label{tab:performance}
    \begin{tabular}{l r r r r r r} \toprule
         Oracle & \# of tests & \makecell[r]{\# of successful\\ queries} & \makecell[r]{\# of unsuccessful\\ queries} & QPT & \makecell[r]{\# of unique\\ query plans} & \makecell[r]{branch\\ coverage}\\\midrule
         % Oracle & \# of tests & \# of successful queries & \# of unsuccessful queries & QPT & \# of unique query plans & branch coverage\\\midrule
         % PQS & 2,981,018k & 3,010,595k & 147,928k & 1.06 & 14,842 & \\
         NoREC & 2,086,646k & 4,207,286k & 149,036k & 2.05 & 172,808 & 63.18\%\\
         % TLP & 1,757,188k & 2,180,736k & 398,919k & 1.47 & 137,743 & \\
         TLP & 976,216k & 2,180,736k & 398,919k & 2.23 & 137,743 & 63.63\%\\
         DQE & 441,350k & 7,502,402k & 21,997k & 17.00 & 486 & 46.71\%\\
         \approach & 497,092k & 1,655,518k & 53,102k & 3.33 & 2,577,603 & 63.06\%\\
         \makecell[l]{\approach \&\\ Expression} & 1,423,068k & 4,411,510k & 326,849k & 3.10 & 7,399 & 63.23\%\\
         \makecell[l]{\approach \&\\ Subquery} & 423,310k & 1,488,817k & 47,141k & 3.51 & 2,755,619 & 62.19\%\\\bottomrule
         % \approach & 68,780k & 228,185k & 6,482k & 3.32 & 708,239 & 63.06\%\\ % old res
         % \approach \& Expression & 1,073,915k & 3,324,371k & 97,491k & 3.10 & 7,378 & 63.23\%\\ % old res
         % \approach \& Subquery & 52,039k & 182,751k & 5,805k & 3.51 & 911,386 & 62.19\%\\\bottomrule % old res
    \end{tabular}
\end{table*}

Table~\ref{tab:performance} shows the results.
On average, \approach has a lower test throughput compared to NoREC and TLP, but a higher test throughput than DQE. Specifically, \approach was approximately 4.20$\times$ slower than NoREC, 1.96$\times$ slower than TLP, and 1.13$\times$ faster than DQE.
Upon further investigation, we have found two reasons why \approach is slower than NoREC and TLP.
Firstly, for each test, \approach required executing at least three queries to derive the test oracle, including original query, folded query, and auxiliary query. 
When using subqueries to create relations, an additional query is required to retrieve the types of the subquery's result, which is used for table creation.
This explains why the QPT exceeded 3 for \approach and its configurations.
Additionally, for \approach, when applying a subquery in an \lstinline[style=codestyle]{INSERT} statement, additional statements are needed to create and drop tables to maintain the database state.
%This accounted for the 3.5 queries per test in \approach with subqueries.
In contrast, NoREC executed only about two queries per test. For TLP, the average number of queries per test was 2.23; the number exceeds 2 as the test oracle randomly either executes the partitioning queries as one query---using \lstinline[style=codestyle]{UNION ALL}---or executes three queries.
For DQE, a test requires not only the three statements \lstinline[style=codestyle]{SELECT}, \lstinline[style=codestyle]{UPDATE}, and \lstinline[style=codestyle]{DELETE}, but also additional statements for two extra columns. These extra columns are used to uniquely identify each row and to track whether a row has been modified.
% For DQE, a test requires at least 3 queries per test; however, DQE generated many unsuccessful queries, meaning that generating a successful test required more queries on average.
Secondly, the execution of queries with subqueries is significantly slower compared to queries with expressions (\ie 7.18$\times$).  
We found that, on average, queries with expressions alone required only 44.73 microseconds for execution, whereas queries with subqueries required 321.19 microseconds.
Although subqueries require more execution time, they cover a significantly greater number of unique query plans, resulting in \approach covering 14.92$\times$ (compared to NoREC) to 5303.71$\times$ (compared to DQE) more unique query plans than other oracles.
NoREC, TLP, and \approach have a similar result of branch coverage, as they use the same statement generator. DQE, however, has lower branch coverage, because it cannot test certain language features, such as \lstinline[style=codestyle]{JOIN}.
However, branch coverage does not effectively illustrate the oracle's effectiveness, as SQLite already achieves 100\% branch coverage with its own test suites, yet still contains logic bugs. Recent research suggests that exercising more unique query plans corresponds to uncovering more interesting and potentially erroneous behaviors in the DBMS under test~\cite{DBLP:conf/icse/BaR23}.

% \begin{table*}[tb]\small
%     \centering
%     \caption{}
%     \label{tab:depthperf}
%     \begin{tabular}{c r r } \toprule
%          The depth of expression & \# of tests & \# of unique query plans \\\midrule
%          1 & 1,151,555k & 7787\\
%          3 & 1,073,915k & 7378\\
%          5 & 865,456k & 5081\\
%          7 & 720,064k & 4246\\
%          9 & 532,295k & 3281\\
%          11 & 381,118k & 2639\\
%          13 & 221,542k & 1735\\
%          15 & 122,031k & 1184
%          \\\bottomrule
%     \end{tabular}
% \end{table*}

\begin{figure}[t]
    \centering
    \begin{tabular}{@{}c@{}}
        \includegraphics[width=.4\linewidth]{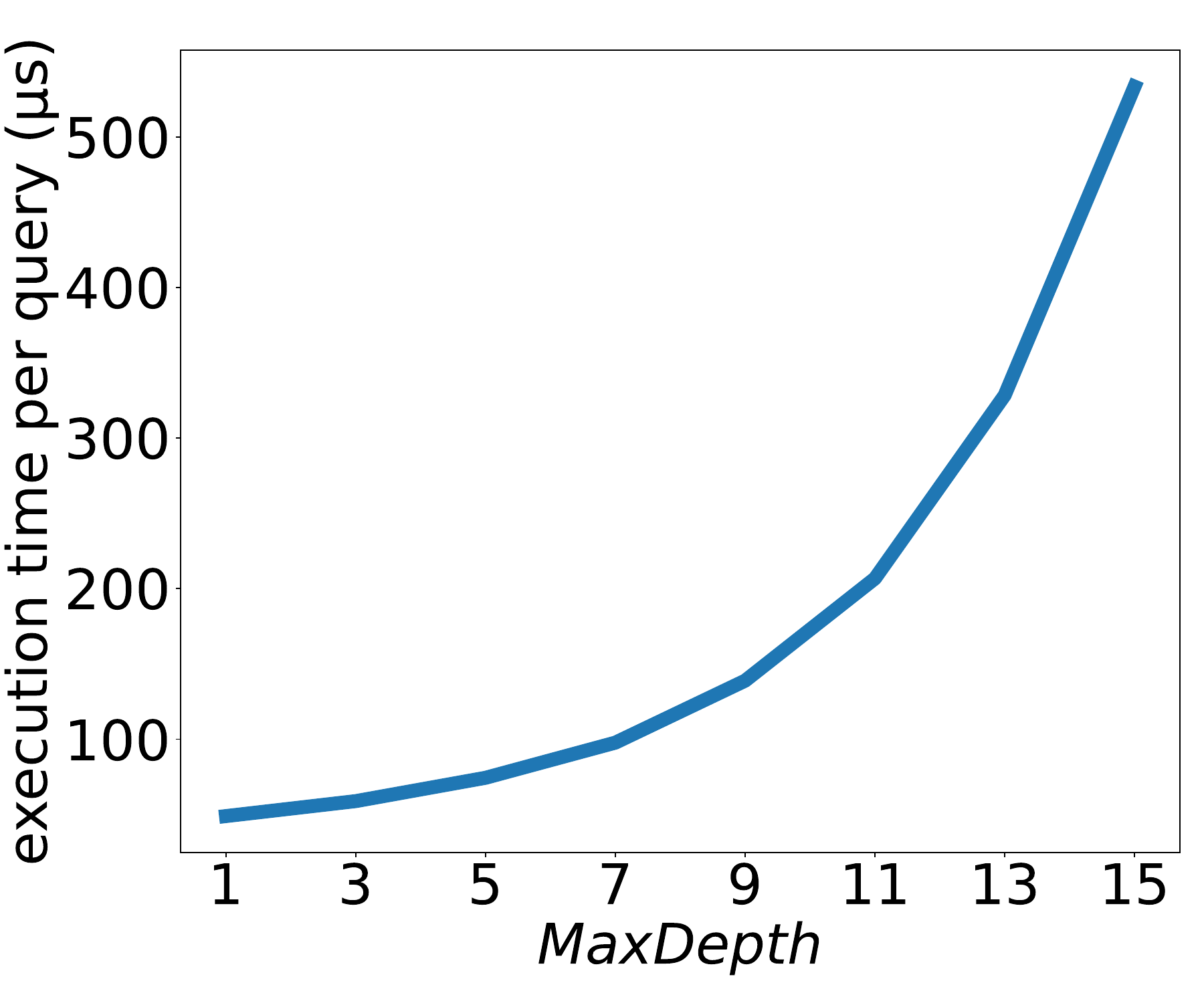}
    \end{tabular}
    \begin{tabular}{@{}c@{}}
        \includegraphics[width=.4\linewidth]{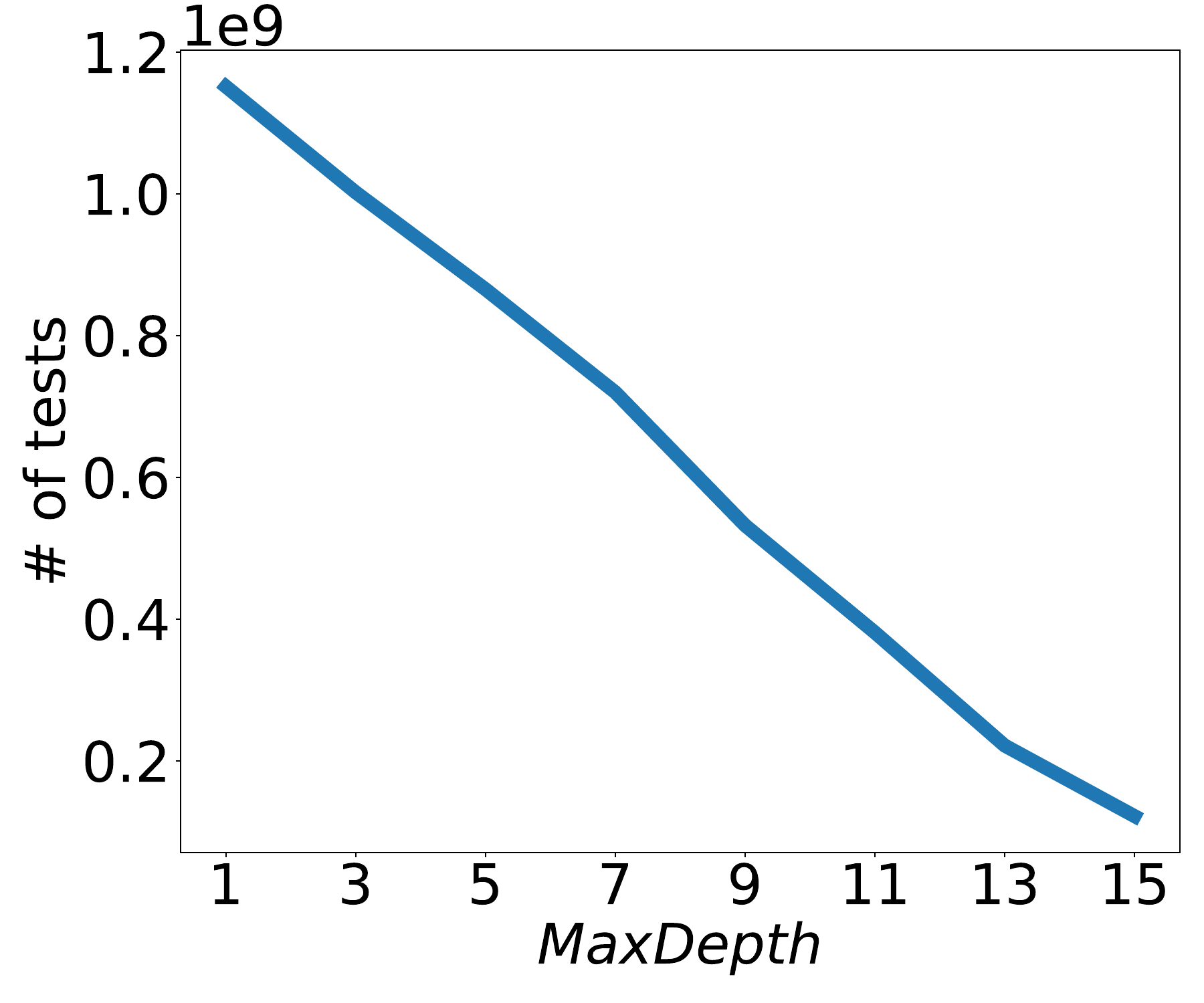}
    \end{tabular}
    % \vspace*{-5pt}
    \caption{The impact of expression complexity on query execution time and test throughput.}
    % \vspace*{-5pt}
    \label{fig:exprdepth1}
\end{figure}

\begin{figure}[t]
    \centering
    \includegraphics[width=.4\linewidth]{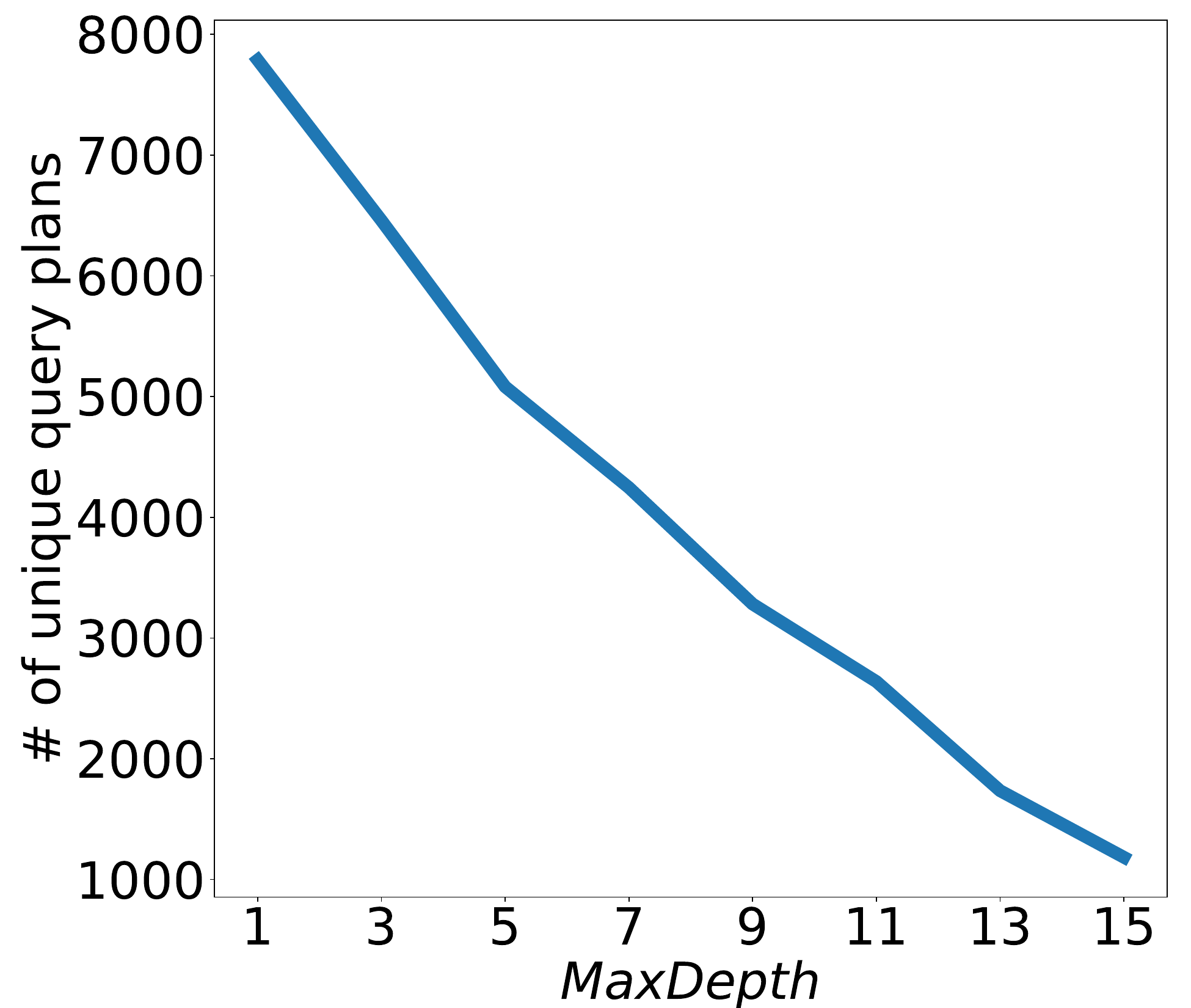}
    % \vspace*{-5pt}
    \caption{The impact of expression complexity on unique query plans.}
    % \vspace*{-5pt}
    \label{fig:exprdepth2}
\end{figure}

\paragraph{Expression complexity.}
We also evaluated the impact of expression complexity on the efficiency of \approach.
To isolate the effect of expression complexity, we conducted our evaluation on \approach \& Expression, excluding the influence of subqueries.
We define the complexity of an expression using \textit{MaxDepth}.
This experiment focuses on examining how complexity influences the average execution time of each query (\ie execution time per query), as well as the overall throughput of \approach (\# of tests).
We conducted this evaluation using the same setup as before (\eg executing the experiment with 10 threads for 24 hours).
The results, as depicted in Figure~\ref{fig:exprdepth1}, reveal how varying of \textit{MaxDepth} influence these performance metrics.
We can observe that, as \textit{MaxDepth} increases from 1 to 15, the average running time for each query increases by 9.91$\times$, and the throughput of our method decreases by 89.4\%.
Therefore, the complexity of expressions significantly impacts the efficiency of our approach, as the DBMSs require more time to execute each query.
However, in our evaluation, we found that most of the bugs we found by the bug-inducing test cases use only shallow expressions after reduction. 
Therefore, during testing, there is no need to generate expressions with great depth.

We also investigated the impact of expression complexity on the diversity of unique query plans (\# of unique query plans) generated when varying \textit{MaxDepth}. 
As shown in Figure~\ref{fig:exprdepth2},
we find that the complexity of the expression has a significant impact on the number of unique query plans, which shows a decreasing trend similar to that of throughput.
We can conclude that increasing expression depth with language features other than subqueries does not significantly exercise additional logic in DBMSs; the use of subqueries in expressions not only increases expression complexity, but also effectively triggers more logic within DBMSs.

\paragraph{Summary.}
\approach generates more unique query plans than existing oracles, suggesting that it exercises interesting functionality in the DBMSs. 
However, this results in slightly lower throughput. Prior research~\cite{DBLP:conf/sigsoft/BohmeF20} suggests that the resources required to find bugs increase exponentially---a constant factor will not significantly decrease the bug-finding effectiveness.
\section{Discussion}
\paragraph{\approach scope}
One potential concern is how \approach can be applied to features not considered in this work. 
First, we believe that any kind of expression can be supported by \approach assuming that it is deterministically computing its result. 
One advanced feature we did not consider are window functions, which can be supported by being used in subqueries. 
In addition, our approach is not specifically designed to test functionality unrelated to expressions, such as clauses (\eg \lstinline[style=codestyle]{LIMIT}). 
While we can apply constant folding and propagation to the predicates of clauses, it is unclear how to replace the clauses themselves. 
Furthermore, like other logic bug detection methods, our approach lacks support for expressions with non-deterministic functions or ambiguous queries.

Our approach primarily applies to the statement types Data Query Language (DQL) and Data Manipulation Language (DML).
It is inapplicable to most statements of Data Definition Language (DDL), Transaction Control Language (TCL), Miscellaneous Language (ML), and Data Control Language (DCL).
% SQL can generally be categorized into six statement types: Data Definition Language (DDL), Data Manipulation Language (DML), Data Query Language (DQL), Transaction Control Language (TCL), Miscellaneous Language (ML), and Data Control Language (DCL). 
% Constant folding and propagation can be directly applied to expressions in the main SQL statement types, including DQL and DML. 
DDL expressions produce results based on data that will be inserted into tables in the future, making \approach inapplicable to constant folding and propagation based on the current database state. 
DCL, TCL, and ML typically do not support the use of expressions as conditions.

\paragraph{The support of closed-source commercial DBMSs}
In our evaluation, we considered only DBMSs whose source code is publicly available. 
One concern could thus be whether our approach could potentially also find bugs in closed-source commercial DBMSs. 
First, multiple of these open-source DBMSs are developed by companies, such as Cockroach Labs for CockroachDB and PingCAP for TiDB. 
Second, we attempted engaging with multiple closed-source commercial DBMS vendors in the past, in whose systems we also found bugs. 
However, we did not receive a response to our bug reports and any potential fixes would be reflected only in subsequent versions, making it difficult to identify their root causes and avoid reporting duplicate bugs.
\section{Related Work}

\paragraph{Detecting logic bugs in DBMSs.}
Several approaches have been proposed to detect logic bugs in DBMSs.
RAGS~\cite{DBLP:conf/vldb/Slutz98} applies differential testing to find logic bugs by comparing the results of a query on different DBMSs or different versions of the same system.
A key challenge that limits its applicability is that SQL dialects differ widely across DBMSs.
Non-optimizing Reference Engine (NoREC)~\cite{rigger2020detecting} and Differential Query Execution (DQE)~\cite{song2023testing} place the same predicate in different clauses, with the expectation that this predicate will retrieve the same rows. 
NoREC primarily focuses on the \lstinline[style=codestyle]{WHERE} clause of \lstinline[style=codestyle]{SELECT} statements, while DQE expands its scope to the \lstinline[style=codestyle]{WHERE} clauses of \lstinline[style=codestyle]{SELECT}, \lstinline[style=codestyle]{UPDATE}, and \lstinline[style=codestyle]{DELETE} statements.
Ternary Logic Partitioning (TLP)~\cite{rigger2020finding} decomposes a query into three partitioning queries, each of which retrieves rows based on the predicates \lstinline[style=codestyle]{p}, \lstinline[style=codestyle]{NOT p}, and \lstinline[style=codestyle]{IS NULL}, respectively. 
TLP  can test the predicate in \lstinline[style=codestyle]{WHERE}, \lstinline[style=codestyle]{GROUP BY}, \lstinline[style=codestyle]{HAVING} clauses, as well as aggregate functions, and \lstinline[style=codestyle]{DISTINCT} queries.
NoREC and DQE rely on predicates by assuming that their results remain consistent regardless of the clause in which they are placed, and TLP leverages that for any given row, exactly one of \lstinline[style=codestyle]{p}, \lstinline[style=codestyle]{NOT p}, and \lstinline[style=codestyle]{p IS NULL} evaluates to true. 
%These three approaches are unable to detect logic bugs in code that is used to process predicates in all clauses, and generates incorrect results for specific expressions, for example, the bug shown in Figure~\ref{fig:illustexample}.
These three approaches are unable to detect logic bugs that affect the evaluation of expressions irrespective of how they are used, making them overlook, for example, the bug shown in Listing~\ref{fig:illustexample}.
%This limitation is also mentioned by DQE in that they cannot detect bugs that consistently affect predicates used in the queries.
Pivoted Query Synthesis (PQS)~\cite{rigger2020testing} generates queries that are guaranteed to retrieve a selected row, based on a naive implementation of operators and functions to be tested, %making it challenging for it to generate and test complex queries.
Transformed Query Synthesis (TQS)~\cite{DBLP:journals/pacmmod/TangWZ0023} generates queries by decomposing a table into multiple sub-tables, to derive a test case and ground truth for queries that join these tables. 
The synthesis methods used by PQS and TQS ensure that the generated query matches a specific value or relationship. This is achieved using their self-built evaluation engine. Therefore, their synthesis techniques have limited support for various language features (\ie TQS only supports equi-join as predicates), as it is challenging to compute the expected results across all of them. This is also why we chose not to compare with PQS and TQS (the latter which is also not publicly available); many language features we test would be difficult to support by these approaches.
A concurrent work called Equivalent Expression Transformation (EET)~\cite{DBLP:conf/osdi/Jiang024} closely relates to our approach. Both \approach and EET operate on expressions and aim to derive queries whose results are the same as the original queries, but the conceptual angle on how they derive them is different. EET introduces tautologies and contradictions while ensuring that the result remains equivalent to the original query.
% determined boolean expressions—constructed using logical operations that always evaluate to either true or false—to add more language features, resulting in an equivalent but more complex query. 
In contrast, \approach applies constant folding and propagation on expressions to replace certain language features with constants, thereby creating an equivalent, but simpler query. Both are black-box approaches, making it difficult to conceptualize which bugs are overlooked by one of the two approaches, but not the other. However, assuming a perfect query optimizer that can simplify any expressions, EET would be ineffective as the unnecessarily complex expressions could be simplified. For \approach, this is not the case, as the query optimizer cannot assume that the database's contents remains the same across queries.

% \paragraph{DBMS test case generation.}
\paragraph{Random and targeted queries.}
Multiple works have improved query generation for DBMSs, which is complementary to our contribution. 
Query Plan Guidance (QPG)~\cite{DBLP:conf/icse/BaR23} mutates the database state to cause the DBMS to guide test case generation towards potentially unseen query plans for subsequent queries. 
SQLRight~\cite{DBLP:conf/uss/LiangL022} mutates the SQL statements based on code coverage feedback to cover more code in DBMSs.
Squirrel~\cite{DBLP:conf/ccs/ZhongC0ZLW20} mutates SQL queries based on an intermediate representation to ensure syntax validity and uses coverage feedback for guidance.
Griffin~\cite{DBLP:conf/kbse/FuLWWJ22} proposes a grammar-free mutation approach for testing DBMSs, using a metadata graph to ensure semantic correctness.

\paragraph{Random and targeted databases.}
Many approaches have been proposed to automate the generation of databases.
Gray et al. proposed approaches for generating billions of database records using multiple techniques~\cite{DBLP:conf/sigmod/GraySEBW94}.
Data Generation Language (DGL)~\cite{DBLP:conf/vldb/BrunoC05} is a domain-specific language designed for generating data that exhibits complex intra- and inter-table correlations.
QAGen~\cite{DBLP:conf/sigmod/BinnigKLO07} is a query-aware database generator designed to produce database states by generating them based on a specified parametric query and set of user-defined constraints, ensuring that the outcomes of the query meet the user requirements. 
ADUSA~\cite{DBLP:conf/kbse/KhalekELK08} leverages a constraint solver to generate database data and the corresponding expected results for a given query and database schema.

% \paragraph{Metamorphic testing.}
% Metamorphic testing~\cite{DBLP:journals/corr/abs-2002-12543} 
% is a methodology for detecting bugs based on so-called metamorphic relations between a test case and its follow-up version.
% The essence of metamorphic testing lies in identifying a metamorphic relation that can be used to generate test cases bound by a specific relation.
% This approach has been successfully applied in various domains~\cite{chen2018metamorphic, segura2016survey}.
% \approach is a metamorphic testing approach, as the generated folded query is equivalent to the original query under the same database state. 
% Compared with current metamorphic testing methods that target logical bugs in DBMS~\cite{rigger2020detecting, rigger2020finding, song2023testing}, \approach adapts the concept of constant optimization in compilers for testing purposes, which we believe is a non-obvious and conceptually interesting perspective.
% In addition, this metamorphic relation has enabled our approach to detect many bugs that remained undiscovered by existing test oracles, as we discussed in Section~\ref{sect:comparison}.

\paragraph{Testing other aspects of DBMSs.}
Besides logic bugs, testing approaches were proposed to find performance issues.
Cardinality Estimation Restriction Testing (CERT)~\cite{CERT} is designed to detect performance issues through the lens of cardinality estimation. For a given query, CERT derives a more restrictive query, for which the cardinality estimator is expected to predict that the query fetches fewer rows than the original query.
APOLLO~\cite{DBLP:journals/pvldb/JungHAKK19} detects performance regression issues in DBMSs by analyzing multiple versions of a given DBMS, and employs a suite of validation checks to minimize false positives.
\section{Conclusion}
In this paper, we have presented a black-box approach for detecting logic bugs in DBMSs, named \textit{Constant-Optimization-Driven Database Testing (CODDTest)}. Our key insight is that, for a given query and fixed database state, constant propagation and folding can be applied to specific expressions of the query, assuming that the result remains unchanged. %, by subsituting the selected expression with constants, and the test oracle relies on validating that the transformed query produces identical results to the original query.
We believe that this idea is non-obvious, as constant propagation and constant folding were originally proposed as compiler optimizations, not for testing DBMSs.
%We have implemented this approach as a tool called \approach.
We have evaluated \approach on five mature and well-tested DBMSs, and found a total of \totalbug bugs. 
Of these, \totallogicbug were logic bugs, and the remaining were internal errors, crashes, and hang-related bugs.
%Our evaluation illustrates that 
%Out of the \totallogicbug logic bugs, 11 were undetectable for the state-of-the-art approaches.
%Of these, 
As indicated by our manual analysis, 11 logic bugs were missed by the state-of-the-art approaches.
%Last, we also analyzed the performance of \approach. 
%While \approach has a lower testing throughput compared to state-of-the-art methods,
\approach generates test cases that exercise more unique query plans, suggesting that it can explore interesting functionality in the DBMS under test. %which has enabled us to find more bugs.
Overall, we believe that \approach is a practical, widely applicable DBMS testing approach that complements existing test oracles for logic bugs.

% \section*{Data-Availability Statement}
% For now, we provide an anonymized version of the artifact at Figshare: \url{https://figshare.com/s/f601c272cebc253bd008}, which includes the implementation of \approach and all experimental results. We will archive this for long-term storage at Zenodo after the acceptance.

\begin{acks}
 This work was partially supported by the National Natural Science Foundation of China under Grant No. 62032010, 62202220. 
 We would like to thank for the financial support from the program of China Scholarships Council (No.202106190065).
\end{acks}

% \clearpage
% \bibliographystyle{plain}
\balance
\bibliography{main}
\end{document}